\documentclass[fleqn,usenatbib]{mnras}

\usepackage[T1]{fontenc}
\usepackage{ae,aecompl}

\usepackage{graphicx}	% Including figure files
\usepackage{amsmath}	% Advanced maths commands
\usepackage{amssymb}	% Extra maths symbols

\usepackage{hyperref}
\usepackage{url}
\usepackage{microtype}
\usepackage{rotating}
\usepackage{booktabs}
\usepackage{threeparttable}
\usepackage{tabularx}

\newcommand{\project}[1]{\textsl{#1}}

\newcommand{\rxte}{\project{RXTE}}
\newcommand{\xmm}{\project{XMM-Newton}}
\newcommand{\rosat}{\project{ROSAT}}
\newcommand{\swift}{\project{Swift}}
\newcommand{\astrosat}{\project{Astrosat}}
\newcommand{\nicer}{\project{NICER}}

\title[The Long-Term Evolution of GRS 1915+105]{Using Machine Learning to Explore the Long-Term Evolution of GRS 1915+105}

\author[D. Huppenkothen et al.]{Daniela Huppenkothen$^{1, 2, 3}$, Lucy M. Heil, David W. Hogg$^{1,2,5}$, Andreas Mueller$^{2}$
\\
   $^{1}$Center for Cosmology and Particle Physics, Department of Physics, New York University, 4 Washington Place, New York, NY 10003, USA \\
  $^{2}$Center for Data Science, New York University, 65 5h Avenue, 7th Floor, New York, NY 10003 \\
 $^{3}$E-mail: daniela.huppenkothen@nyu.edu \\
  $^{5}$Max-Planck-Institut f\"{u}r Astronomie, Heidelberg, Germany \\
  $^{6}$Simons Center for Data Analysis, 160 Fifth Avenue, 7th floor, New York, NY 10010, USA
}

\date{Accepted XXX. Received YYY; in original form ZZZ}

\pubyear{2016}

\begin{document}
\label{firstpage}
\pagerange{\pageref{firstpage}--\pageref{lastpage}}
\maketitle

\begin{abstract}
Among the population of known galactic black hole X-ray binaries, GRS 1915+105 stands out in multiple ways. It has been in continuous outburst since 1992, and has shown a wide range of different states that can be distinguished by their timing and spectral properties. These states, also observed in IGR J17091-3624, have in the past been linked to accretion dynamics. %However, currently no clear physical picture exists of how the ensemble of states is produced and what physical quantities drive them. 
Here, we present the first comprehensive study into the long-term evolution of GRS 1915+105, using the entire data set observed with \rxte\ over its sixteen-year lifetime. We develop a set of descriptive features allowing for automatic separation of states, and show that supervised machine learning in the form of logistic regression and random forests can be used to efficiently classify the entire data set. For the first time, we explore the duty cycle and time evolution of states over the entire sixteen-year time span, and find that the temporal distribution of states has likely changed over the span of the observations. We connect the machine classification with physical interpretations of the phenomenology in terms of chaotic and stochastic processes.
\end{abstract}


%%%%%%%%%%%%%%%%%%%%%%%%%%%%%%%%%%%%%%%%%%%%%%%%%%

%%%%%%%%%%%%%%%%% BODY OF PAPER %%%%%%%%%%%%%%%%%%

\section{Introduction}
Black hole X-ray binaries (BHXRBs), systems containing a stellar-mass black hole and a main-sequence companion, are some of the best test cases of fundamental physics, including tests of general relativity in strong gravity, plasma physics in accretion discs and particle acceleration in astrophysical jets. 
Due to the relative simplicity of black hole mass scaling, they may also be seen as smaller analogues to their super-massive counterparts in Active Galactive Nuclei (AGN), by providing a window into physical processes on much shorter time scales and at much higher observable fluxes.

Among the known BHXRBs, GRS 1915+105 holds a special position. Discovered as a bright, $0.35$ Crab X-ray source \citep{castrotirado1994} with the WATCH all-sky monitor on the GRANAT space telescope \citep{castrotirado1992}, it also became known as the first galactic source known to exhibit superluminal jets \citep{mirabel1994, fender1999} and was hence termed a `microquasar' for its similarities to its supermassive counterparts. 
Despite being highly absorbed, optical identification of a K-M III type non-degenerate companion with the Very Large Telescope allowed a mass estimate of $14\pm 4\,M_\odot$ \citep{greiner2001}, recently revised via trigonometric parallax to a slightly lower mass of $12.4^{+2.0}_{-1.8}\, M_\odot$ and a distance of $8.6^{+2.0}_{-1.6}\,\mathrm{kpc}$ \citep{reid2014}. 
Since its discovery in 1994, GRS 1915+105 has been monitored repeatedly with instruments across all wavelengths, providing the first solid evidence of a coupling between accretion disc and jet: hard X-ray dips in the complex light curves of GRS 1915+105 were found to be associated with bright events at infrared and radio wavelengths \citep{pooley1997, eikenberry1998a, eikenberry1998b, kleinwolt2002}. Additionally, steady jets seem to be present during periods of prolonged hard X-ray emission \citep{foster1996, dhawan2000, fuchs2003}. 

What sets GRS 1915+105 apart from the remaining sources in the sample of known BHXRBs is its X-ray variability. Variability in both flux and spectrum is expected from these sources since their accretion disc likely undergoes turbulence driven by magnetic instabilities. However, GRS 1915+105 is known to exhibit complex X-ray light curves spanning at least 14 different patterns \citep{belloni2000, kleinwolt2002, hannikainen2003, hannikainen2005}. These complex patterns are known to repeat almost identically, sometimes with months to years between occurrences. It was thought to be unique in its behaviour until the detection of a second source, IGR J17091-3624 \citep{altamirano2011}, exhibiting similar variability. 
The variability, going hand-in-hand with spectral changes on short time-scales, is difficult to explain with standard accretion theory. Yet understanding the origin and formation of these patterns is crucial, as they are clearly not random and encode information about the accretion disc. \citet{belloni1997a, belloni1997b, belloni2000} suggested that all variability patterns observed in GRS 1915+105 decompose into three basic states, termed A, B and C, based on spectral and variability characteristics. These three fundamental states seem to roughly correspond to similar spectral and variability properties in other BHXRBs, in particular to the low-hard state with a hard spectrum and the presence of strong variability (LHS; state C in GRS 1915+105) and the very high state with a soft spectrum and little variability (VHS; state B at high flux and A with similar spectrum, but lower average flux).
%AM can you describe A, B and C in some way? Were they done for IGR J17091-3624 or GRS 1915+105 or both? It's not entirely clear from the text [DONE]

While \citet{belloni2000} point out that their state classification is mainly intended for easy categorization of observations, it is clear that the observed variability patterns are intimately linked to the underlying accretion physics. \citet{naik2002} observed that certain variability classes ($\alpha$ and $\rho$ in the \citealt{belloni2000} classification scheme) are preferably observed before and after prolonged intervals of the source in a type-C state with a hard spectrum, indicating that there exists a connection between the states as classified by \citet{belloni2000} and the long-term behaviour of the source, which may possibly be linked to mass accretion rate. If this is the case, then the complex variability leads to interesting prospects for studying accretion disc dynamics at high mass accretion rates. 

Based on a similar idea, \citet{misra2004, misra2006} grouped the original 12 classes into three groups based on an analysis of the correlation dimension, a proxy for distinguishing stochastic from chaotic processes. They found representatives of both chaotic and stochastic processes (see also \citealt{harikrishnan2011} for follow-up work), with five of the original classes showing non-linear deterministic (i.e.\ chaotic) behaviour ($\theta$, $\rho$, $\alpha$, $\nu$, $\delta$), three exhibiting purely stochastic behaviour ($\phi$, $\gamma$, $\chi$) and four showing a mix of chaotic and stochastic behaviour ($\beta$, $\lambda$, $\kappa$, $\mu$). The results were recently confirmed by \citet{sukova2016} using recurrence analysis and indicate a complex interplay between the governing physical properties---e.g.\ mass accretion rate and viscosity---and the observable X-ray emission.
% AM "individual groups" as opposed to what kind of groups? FIXED
% AM "both possibilities" -> be explicit "Found representations of both chaotic and stochastic processes" or something like that FIXED

On the other hand, \citet{polyakov2012} looked at the stochastic variability in all thirteen classes characterized in \citet{belloni2000} and \citet{kleinwolt2002} using Flicker Noise Spectroscopy and found four different modes of stochastic behaviour, which they connected to viscosity fluctuations in the accretion disc. Their results broadly agree with those of \citet{misra2006}, though they point out that for some observations, the quality of the data does not allow a firm identification of the variability with the proposed modes.

It is likely that the complex, recurring variability patterns are driven by global instabilities in the accretion disc, i.e.\ non-linear, deterministic processes governed by the global dynamical evolution of the accretion disc and driven by a few global parameters, for example the accretion rate. \citet{massaro2014} show that the striking patterns observed in the $\rho$ state, also named `heartbeat` state for its quasi-periodic pulses, can be described by a limit cycle caused by a fairly simple system of non-linear ordinary differential equation. Their model indicates that the burst recurrence time largely depends on a parameter steering the forcing in the system, and suggest that either variations in the mass accretion rate or viscosity may act as the driving force behind the observed oscillations in this state, in line with hydrodynamic simulations \citep{nayakshin2000, merloni2006} and detailed observations of spectral changes \citep{neilsen2011, neilsen2012}.

It is clear that the state changes in GRS1915+105 must in some way depend on global properties of the accretion disc, and can act as probes of physical processes within the disc as well as the coupling between the disc and the jet. Thus, understanding the properties of these states and the long-term evolution of GRS 1915+105 is of crucial importance. However, studies to date largely concentrate on either individual states or subsets of the available data based on the previous classification of the first four years of \rxte\ data. 
The purpose of this paper is a study of the full 16-year data set of GRS1915+105 observed with the Proportional Counter Array (PCA) onboard the \textit{Rossi X-ray Timing Explorer} (\rxte). We choose a machine learning approach, novel in this context, to characterize and classify the states in GRS 1915+105. 

Machine learning is a sub-field of computer science concerned with learning patterns from data. In recent years, it has been employed very successfully in a range of different sciences (for an introduction, see e.g. \citealt{bishop2006} or \citealt{ivezic2014} for an astronomy-focused textbook). Machine learning as relevant for astronomy can be broadly separated into two types. In supervised machine learning (either classification or regression), a training data set is available for which the outcomes are known. This requires that such a previous data set exists for which the labels (or regression variables) are known from either human classification or other methods. 
Unsupervised machine learning, conversely, does not assume that the desired output (e.g.\ labels in classification or continuous variables in regression) is known, but aims to actively learn it from the data itself, subject to some assumptions and constraints that depend on the precise method used. In astronomy, machine learning has recently been used in a large variety of contexts, including among many others the estimation of photometric redshifts in the Sloan Digital Sky Survey \citep{carliles2010, beck2016}, automatic classification of galaxies using training data from the Galaxy Zoo project \citep{banerji2010, dieleman2015}, variable X-ray source classification for \rosat \citep{mcglynn2004} and \xmm\ \citep{farrell2015}, modeling the \swift/BAT trigger algorithm \citep{graff2015}, and distinguishing long and short Gamma-Ray Bursts \citep{tarnopolski2015}.
The existing \rxte\ data set for GRS 1915+105 is particularly well suited for a machine learning approach: the source was subject of one of the most comprehensive X-ray monitoring campaigns performed with \rxte, yielding a data set of sufficient size for automatic classification while being too large to be classified by hand in its entirety. It shows fourteen discrete classes, and a fraction of the data set has been annotated by hand in the past, yielding the training set required for supervised machine learning tasks.

\begin{figure*}
\begin{center}
\includegraphics[width=\textwidth]{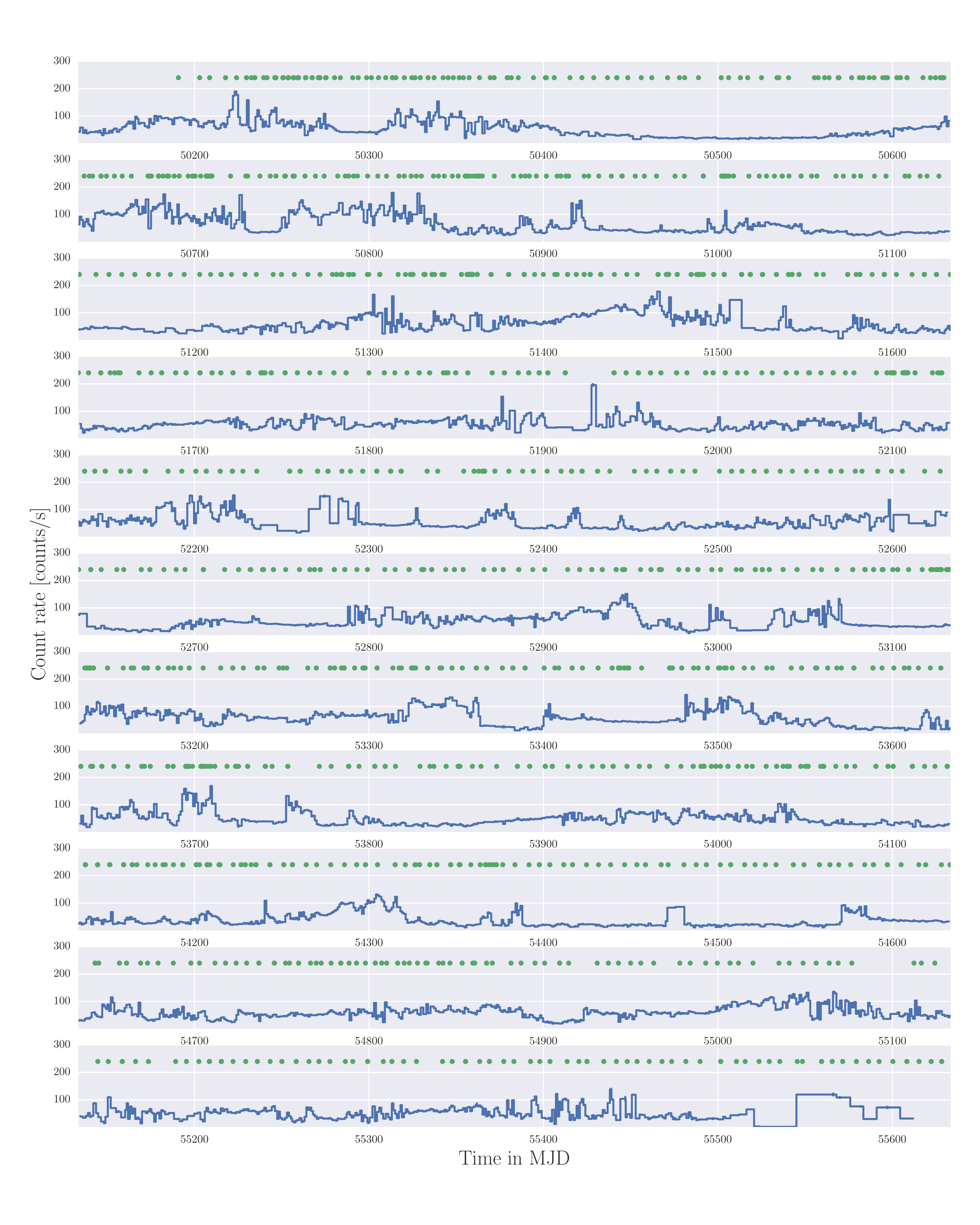}
\caption{\rxte\ All-Sky Monitor (ASM) light curve in Modified Julian Date (MJD) for the entire duration of the \rxte\ mission. Each panel covers $500$ days. The solid blue line is the ASM light curve. The green dots represent the start points of the \rxte/PCA observations with high enough time resolution to be relevant for this analysis. The Figure shows that the \rxte/PCA observations span the entire lifetime and provide an approximately regular sub-sample with high coverage in time, though each observation is short.}
\label{fig:asm_total}
\end{center}
\end{figure*}

In this paper, we show that efficient classification using machine learning can be done, and present ways in which it can be used to infer the physical properties of the source. In Section \ref{sec:observations}, we introduce the data set and the pre-processing performed. Because few machine learning algorithms perform well on raw data, we explain how we constructed \textit{features}---summary statistics of the raw light curves that allow the algorithm to distinguish between classes---in Section \ref{sec:featureengineering}. In Section \ref{sec:supervised}, we present the results of the supervised classification, while in Section \ref{sec:discussion}, we put our 
results in the broader context, discuss limitations of the current approach and show potential avenues for future research.

\section{Observations and Data Preparation}
\label{sec:observations}

We used all available \rxte\ observations of GRS 1915+105 between 1996 and 2011 with data in GoodXenon or EventMode ($1712$ observations). Light curves were extracted with a resolution of $0.125\,\mathrm{s}$ in $4$ energy bands: $W = 3 - 75$ keV, $L = 3 - 6$ keV, $M = 9 - 15$ keV, and $H = 15 - 75$ keV. While the energy ranges will not be exactly the same from light curve to light curve due to different detector modes as well as gradual changes in the sensitivity of individual channels over time, channels were included or excluded as necessary to keep the energy ranges as constant as possible. Out of a total of $1712$ observations, $20$ have no high-band data and are thus excluded, for a total of $1692$ observations included in the analysis (see Figure \ref{fig:asm_total} for the long-term light curve observed with \rxte's All Sky Monitor (ASM), with the locations of pointed PCA observations marked). 
In the following, we use the $3 - 75$ keV band for all time series and power spectral features, and form two hardness ratios that encode energy spectral changes within and between states. Hardness ratio 1 (HR1) is defined as $\mathrm{HR}1 = M/L$ (mid-energy band divided by low-energy band) and hardness ratio 2 (HR2) as $\mathrm{HR}2 = H/L$ (high-energy band divided by low-energy band) to capture spectral changes in a model-independent way.

%{\bf Lucy: Please check whether the paragraph above is correct or whether there is information missing? LH: This looks fine to me - from memory the 75 keV cut-off was due to some Event mode LCs not having the highest band available. It's probably not worth mentioning directly in the paper but if the referee queries it I will double check.}

\begin{figure}
\begin{center}
\includegraphics[width=9cm]{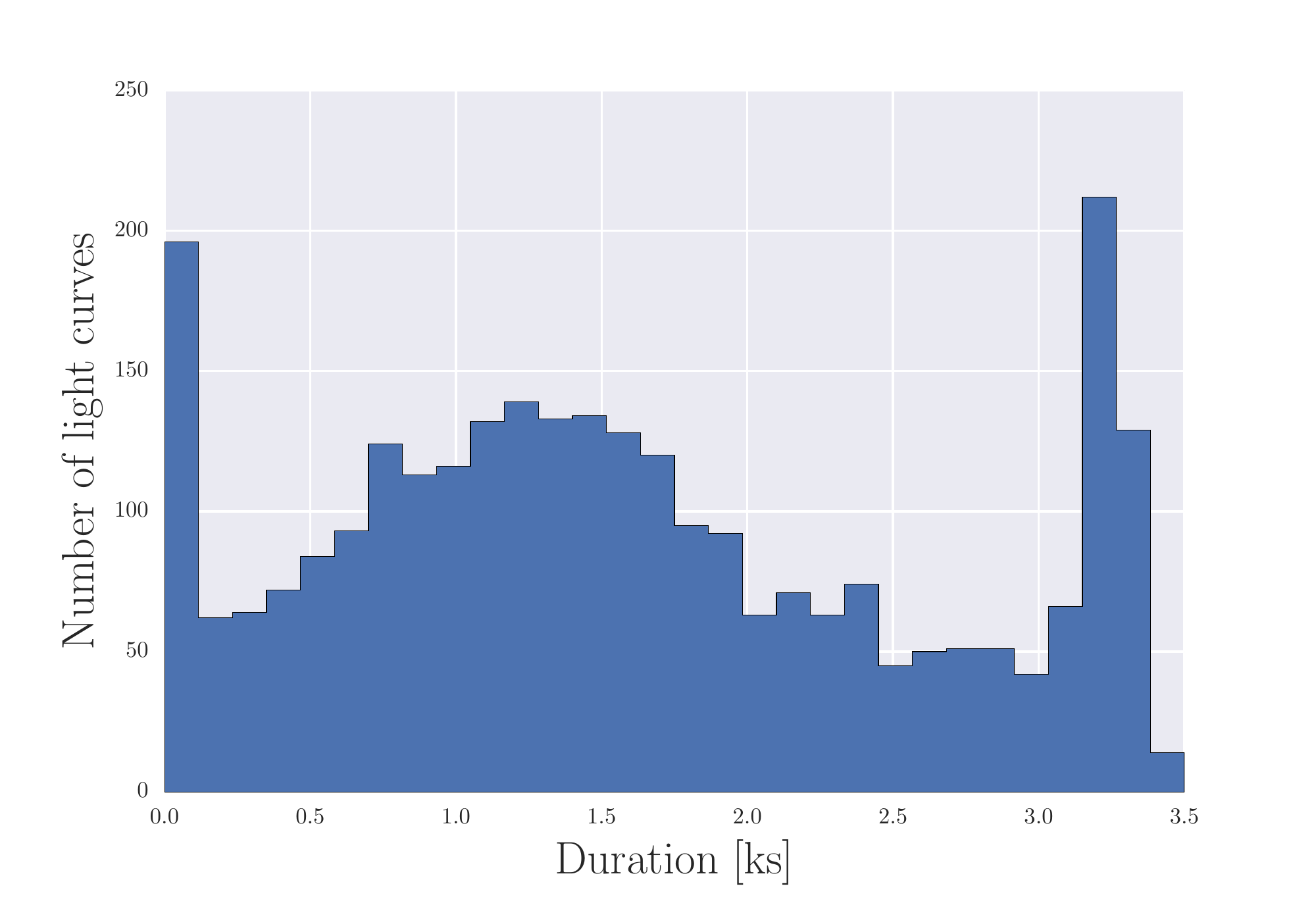}
\caption{Histogram of the durations of all observations used in the analysis. Most observations have durations of $500$ --- $3500$ seconds. Note that these reflect total durations for a given observation without application of Good Time Intervals (GTIs); in the analysis below, these durations may be shortened or split in parts by detector failures and the $90$-minute orbit of the space craft.}
\label{fig:obsdurations}
\end{center}
\end{figure}

Figure \ref{fig:obsdurations} shows a histogram of the durations of individual observations. Most observations have a duration of $\sim\!2000 \,\mathrm{s}$, with only a small subset being significantly longer.
In practice, many light curves are shorter, since data drop-outs and interruptions in the observations lead to good time intervals that are shorter than the nominal observation time. This is an important limitation to keep in mind, given that many of the patterns observed in the light curves of GRS 1915+105 tend to be of the order of $\sim\! 1000 \,\mathrm{s}$ long.  This also leaves us with an important decision to make: do we use all segments regardless of length, or do we produce light curves of equal length for the classification, at the risk of loosing the shortest light curves? The latter is preferable in order to avoid systematic biases in our features (which, in the case of summary statistics, might depend on the number of data points in the light curve) and because some features are structured such that light curves of different duration give feature vectors of different lengths, making the later classification task vastly more complex. 

This implies that there is a trade-off between descriptiveness and sample completeness: when choosing long segments, we likely encapsulate more of the characteristic behaviour of a state, which can sometimes consist of cycles lasting more than a thousand seconds. On the other hand, if we choose long segments, we necessarily exclude all light curves that are shorter than that, for example because their Good Time Intervals (GTIs) only allowed for shorter segments. Here, we pick a segment length of $1024\,\mathrm{s}$ as a reasonable trade-off between being descriptive (generally, the patterns observed in \citet{belloni2000} last $\sim\!1000\,\mathrm{s}$ or so) and providing sufficient samples for classification. Note that we also choose overlapping segments starting every $256\,\mathrm{s}$, both for data augmentation as well as to account for phase shifts in periodic patterns. This leaves us with a total of $8506$ data segments of $1024\,\mathrm{s}$ duration, each of which consists of $8192$ data points in each of the four energy bands. Because of the overlap between them, segments derived from the same contiguous observation interval will not be independent. This is a standard data augmentation practice in machine learning applications and does not affect our conclusions as long as samples in the validation and test set are independent from the training data set, which we ensure as described in Section \ref{sec:featureselection} below.

\section{Feature Engineering}
\label{sec:featureengineering}

While some machine learning algorithms can produce reliable classifications on raw data (e.g. a light curve), we find that these algorithms fail on the problem at hand for a number of reasons. The data set to be explored here is relatively small, in machine learning terms, with some classes having less than ten examples in the set of examples with human annotations. Additionally, the light curves show complex periodic patterns whose phases are random with respect to the start of an observation. Thus, different light curves of the same class are phase-shifted and may appear as very different to a machine learning algorithm purely due to this phase shift alone.

Instead, we reduce the number of dimensions by extracting \textit{features}, descriptive summaries of the raw data that will allow for efficient separation of the various classes in feature-space.
In the following, a \textit{sample} is a single instance of the ensemble to be classified, in our case an RXTE data segment (consisting of a light curve in four energy bands) of GRS1915+105, i.e.\ a $4 \cdot 8192 = 32768$-dimensional vector. For each sample, we compute a set of features for classification. 

Feature engineering is the most important and most difficult part of any machine learning problem. It is here where domain knowledge of the problem at hand becomes crucial to finding the most informative features to be used by the computer in the subsequent classification task. 
We use the previous (human-based) annotations by \citet{belloni2000}, supplemented with additional annotations published in \citet{kleinwolt2002} and \citet{hannikainen2003} to guide the feature engineering task. With relatively high-resolution light curves ($\Delta t = 0.125 \,\mathrm{s}$) in four energy bands, there is a multitude of possible features in time, energy and frequency domains that could potentially inform our choices. Because we aim to automate the manual classification in \citet{belloni2000}, we base our feature engineering on similar features such as the hardness ratios and overall appearance of the light curves. We also supplement the feature set derived from the time series and hardness ratios with properties of the power spectrum.

\subsection{Time Series Features}

Because it is difficult to encapsulate the large variety of shapes observed in the light curves of GRS 1915+105, we use a mix of very simple summary features and extract a number of features from an autoregressive model (as explained below). The summary features are: the mean count rate, median count rate, total variance, skewedness and kurtosis in the light curve segment in the $3 - 75$ keV band. 

The light curves observed from GRS1915+105 show a very rich variability behaviour, including complex patterns not well represented by the summary features listed above. Encapsulating these complex variability patterns in a few parameters is generally difficult: for example, any representation must be phase shift-invariant. That is, for roughly periodic patterns, features should look very similar regardless of where in the cycle a light curve begins. We attempt to encapsulate the variability in a simple autoregressive model, where the data $y_t$ at any given point in the light curve $t$ depends on a linear combination of $k$ data points immediately before:
% AM "extract a number of features from a linear model." I feel this deserves more of an explanation here. Given the amount of ML knowledge you assume of the reader, this is hard to understand (and even I can't say for sure what you mean without reading on) [after reading on, I find that none of the interpretations I had was the one I imagined, you might want to say "using an autoregressive model, as explained below"]
% AM this kind of model is called autoregressive.Not an expert in these but I think you should mention the name.

\begin{equation}
y_{t} = c + \sum_{i=1}^k{\left( w_i y_{t-i}\right)} ,
\end{equation}
% AM is there no bias term? YUP
% AM did you do this yourself or use sklearn ridge regression? SKLEARN RIDGE REGRESSION

\noindent  where the $k$ elements $w_i$ of vector $w$ specify the weights, and we define a vector of all $k$ relevant previous measurements $X_t = y_{t-k:t}$ for use below. The weights encode the relative importance of previous $k$ data points on point $y_{t}$. Because these weights should be different for different classes, we expect them to be useful summaries of the complex temporal structure encoded in the light curves.

We minimize the following equation with respect to the weight vector $w$ to infer the optimal weights:

\begin{equation}
\min_w ||\langle w, X \rangle - y||^2 + \lambda ||w||^2 \; ,
\end{equation}

\noindent where $\lambda$ is a regularization parameter that controls for overfitting of the data. We run this optimization for each segment independently, and extract the weight vectors as features to be used in the subsequent classification.

The parameter $k$ defining the number of data points relevant in determining the data point $y_{t+1}$ and consequently the number of weights is a free parameter to be estimated. The final free parameter is the temporal resolution $\Delta t$ of the light curves. In principle, it is possible to run the feature extraction on the unbinned light curves with a resolution of $\Delta t = 0.125\,\mathrm{s}$. However, averaging a set of $n$ neighbouring bins may reduce variance due to measurement noise and thus lead to cleaner features. The parameter space for these features was explored via the validation set and will be explained in more detail in Section \ref{sec:freeparams}.

\begin{table*}[hbtp]
\renewcommand{\arraystretch}{1.3}
\footnotesize
\caption{Model Parameters}
\begin{threeparttable} 
\begin{tabularx}{\textwidth}{p{2.0cm}p{2.0cm}p{5.0cm}p{1.0cm}p{6.0cm}}
\toprule
\bf{Feature Set} & \bf{Parameter} & \bf{Meaning} & Best Value &  \bf{Possible Values} \\ \midrule
		& C & Regularization magnitude & $1$ & $[10^{-3}, 10^{-2}, 0.1, 1, 10, 100]$ \\ \midrule
 Autoregressive Model & $\Delta t$ & Light curve time resolution & $0.125$ & $[0.125, 1.0, 2.0, 6.25]$ \\
		& $k$ & Number of time bins determining current time bin & $10$ & $[2, 5, 7, 10, 20, 30, 50, 80]$ \\
		& $\lambda$ & Regularization parameter & $1000$ & $[0.01, 0.1, 1, 10, 20, 50, 100, 1000]$ \\ \midrule
Power spectrum PCA & $N$ & Number of components & $10$ & $[1,2,3,5,10,20,50,100]$ \\

 \bottomrule
\end{tabularx}
   \begin{tablenotes}
      \item{}
\end{tablenotes}
\end{threeparttable}
\label{table:parameters}
\end{table*}

\subsection{Power Spectral Features}

We use power spectral features based on the power colours defined in \citep{heil2015}. We compute power spectra in fractional rms normalization for all available light curves and integrate over frequencies in order to compute the fractional rms amplitude in different frequency bands. 
Following the power colours defined in \citet{heil2015}, we choose our bands to be $P_\mathrm{A} = 0.0039-0.031 \,\mathrm{Hz}$, 
$P_\mathrm{B} = 0.031-0.25 \,\mathrm{Hz}$, $P_\mathrm{C} =  0.25-2.0 \,\mathrm{Hz}$ and $P_\mathrm{D} = 2.0-16.0 \,\mathrm{Hz}$. We also construct power colours $\mathrm{PC}_1 = P_\mathrm{C}/P_\mathrm{A}$ and  $\mathrm{PC}_2 = P_\mathrm{B}/P_\mathrm{D}$.
In some states, a quasi-periodic oscillation (QPO) is clearly present. As a simple proxy and to avoid time-consuming power spectral fitting, we design a feature composed of the frequency where $\nu P_\nu$, i.e. each power spectral bin multiplied by its frequency,  has its maximum. This feature generally encodes the frequency that dominates the overall variance. If a QPO is present, this feature will encode the frequency of that QPO. For states without QPO, the maximum is generally lower and set by the broadband noise component.

Because these features offer only an incomplete description of the power spectrum in different states, in particular the presence of a QPO cannot be completely described by the prescriptions above (partly because the power spectral bands are much broader, thus a QPO might not have a pronounced effect). Instead, we build a Principal Component Analysis (PCA; \citealt{pearson1901}) representation of the power spectra and include the principal components as features. The number of PCA components, $N_\mathrm{PCA}$ is a free parameter, and will be discussed in more detail in Section \ref{sec:freeparams} below. 

Note that because the QPO moves in frequency, it is in principle possible for its centroid frequency to fall on the edge of a power spectral band. If the QPO frequency moves within a state, this adds some variance to the power colour measurements of samples in that state, as the resulting power colours will lie roughly in between those that would be derived if the QPO were completely contained in either of the adjacent bands. We believe that this effect is very rare in practice, since (1) as we show in Section \ref{sec:featureselection} below, $\mathrm{PC}_1$ is our most descriptive feature and can separate more than half of the validation data set into its correct classes by itself; (2) it is only one of several power spectral features used in the classification, thus the classifier has additional information to draw on when classifying a given sample.

\subsection{Hardness Ratio Features}

\citet{belloni2000} showed that the different classes occupy different positions in the space spanned by HR1 and HR2. While for most classes, there seems to be a strong (approximately linear) correlation between HR1 and HR2, some classes show more complex correlations where the source follows curved tracks through this space. In order to characterize the properties of the spectral evolution, we extract mean, skewedness and kurtosis from each hardness ratio separately. Additionally, we extract the covariance matrix of HR1 and HR2 for each segment, corresponding to the variance of each hardness ratio as well as the covariance between them, yielding a total of $9$ features based on the spectral evolution. It is worth noting that we explored the use of other techniques to extract hardness ratio features, notably 2D histogram maps, PCA and manifold learning techniques, and found them to be no better than the summary statistics above, thus we chose the latter for their simplicity and straightforward interpretability.

\subsection{Feature Selection}
\label{sec:featureselection}

We randomly split the observations in training, validation and test data sets, with $50\%$ of observations in the training set and $25\%$ of all observations in the validation and test sets, respectively. This results in $4668$ samples in the training data set, $2094$ samples in the validation set, and $2450$ sample in the test data set. The differences in samples in the validation and test set are due to the fact that we split \textit{observations} (i.e. before creating segments of equal length) rather than samples. This is necessary because we extract overlapping segments, thus picking randomly from segments would lead to the loss of independence between training, validation and test data sets. As Figure \ref{fig:asm_total} shows, individual observations are generally separated in time, and can thus be considered independent.

For feature selection and supervised classification, we use the combined previous classifications by \citet{belloni2000}, \citet{kleinwolt2002} and \citet{hannikainen2003} in order to capture all $14$ currently known states, but include only classifications where the entire observation was seen to be in a single state in order to avoid accidental mis-classification as the source switches states within an observation. This yields a total of $1884$ previously classified samples, with $885$ classified samples in the training set, $480$ samples in the validation set and $519$ samples in the test set, respectively. Note that the data set is heavily imbalanced with respect to class representation: previously, the source was known to spend the majority of its time in the $\chi$ state, while other states (e.g.\ $\eta$ and $\omega$) were only seen in one or two observations. 

Initial visualization showed that some features span a wide range of values. Because some machine learning methods, specifically the logistic regression model introduced below, tend to do better in well-behaved (linear) feature spaces, we take the logarithm of features that extend over several orders of magnitude and use the validation set to confirm that this improves classification with the logistic regression model. In particular, these features are: the variance of the light curve, the frequency where $\nu P_\nu$ has its maximum, all fractional rms values and power colours derived from the power spectrum, and the variances of both hardness ratios. 

\subsubsection{Free Parameters}
\label{sec:freeparams}

To estimate the free parameters of the model (see Table \ref{table:parameters} for an overview) we followed the procedure outlined above: we split the data set with human annotations into training and validation sets. We then used the former to train the algorithm, and the latter to test performance and find the combination of parameters that maximizes performance.

\begin{figure}
\begin{center}
\includegraphics[width=9cm]{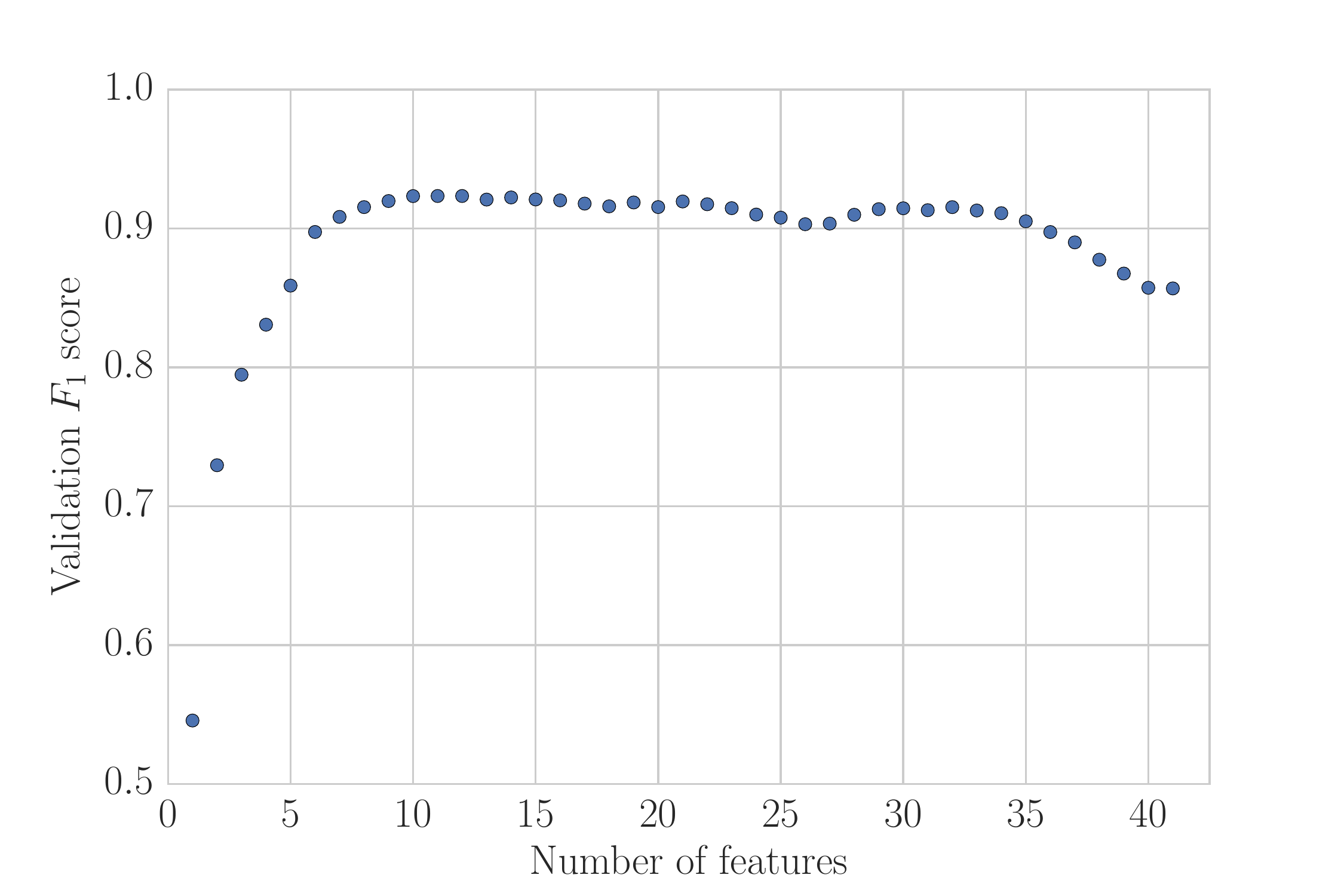}
\caption{The greedy search for the most important features: the number of features used for classification versus the accuracy (fraction of correctly classified samples) of classification in each case. The search shows that a simple logistic regression approach can yield an $F_1$ score of  $>91\%$, and that $10$ features seem to be largely sufficient to classify data from GRS 1915+105. Adding higher features does not add any improvement in predictive power, and can in some cases be detrimental to performance, if it causes the logistic regression model to overfit.}
\label{fig:scores}
\end{center}
\end{figure}

In order to estimate which parameters will yield the optimal results, we use supervised learning in the form of logistic regression (\citealt{cox1958}; as implemented in \textit{scikit-learn}; \citealt{scikit-learn}). Logistic regression is one of the simplest classification algorithms. It defines a linear model analogously to linear regression, but because outcomes are discrete rather than continuous, it uses a binomial distribution (multinomial distribution if more than two outcomes are possible) instead of a normal distribution in defining the likelihood \citep{cox1958}. In practical terms, logistic regression aims to draw a hyperplane in the $N$-dimensional space spanned by all features such that the hyperplane separates samples belonging to a given class in the training set from the remainder. Multi-class classification is performed either using a multinomial distribution or by using a one-versus-all scheme: for each class in the training set, a separate hyperplane is drawn such that the split between samples belonging to said class and the remaining samples is maximized. We use the the least squares (L2) norm for regularization, which introduces an additional parameter, $C$. This parameter is used to balance the ability of the model to produce accurate predictions against tendencies to overfit. It is worth noting here that we also attempted the supervised machine learning task with other algorithms that use different strategies for finding decision boundaries between classes, most notably linear support vector machines \citep{guyon1993,cortes1995} and random forests \citep{breiman2001}, and found no improved performance compared to the logistic regression classifier, thus we keep the latter in the following for its interpretability.

For each combination of parameters listed in Table \ref{table:parameters}, we use the training set to train the model with these parameters and test the performance with the validation set. For performance assessment, we use the $F_1$ score, a harmonic mean of \textit{precision} and \textit{recall}. In binary classification, precision refers to positive predictive value, i.e.\ the number of true positive examples of a class divided by all positive classifications. Recall, in turn, refers to the true positive rate, the ratio of true positives to all correctly identified samples. In the multi-class case considered here, the $F_1$ score is computed for each class in turn and averaged for all classes. We choose a weighted average of the $F_1$ score to account for the class imbalances in terms of samples.
We do use accuracy when assessing performance of the classification with the final, trained model in Section \ref{sec:supervised} below, since it can be straightforwardly interpreted as the fraction of samples the classifier identified correctly.

Using the approach described above, we arrive at the best values used for the classification, yielding a total of $41$ features for each of the $8506$ data segments. This comprises the features explicitly named above, as well as $10$ weights from the autoregressive model, and $10$ components from performing PCA on the power spectra.

\subsubsection{Feature Importance}

In order to assess the relative predictive power of each feature, we implemented a greedy search to find the most important features in our set. 
Once more, we used supervised learning with the previously-classified labels as the training set, but computed the validation score for each feature independently, as if it was the only feature available for classification. We set the feature with the highest validation score as the most important feature, and then perform a second pass, this time using the combination of the winner of the first round with every other feature. Again, we picked the combination with the highest score, and continued this process until all features were exhausted. This procedure answers simultaneously two questions. (i) It provides a ranking for the relative predictive power of each feature, and (ii) it provides an assessment whether all features are required to classify the data. The latter need not necessarily be true: some features (e.g. the power colours) are combinations of other features, and thus all features might not be required.

In Figure \ref{fig:scores}, we show the results of the greedy search. We find that the $F_1$ score is generally high: $\sim\!\! 91\%$ on the validation set for the the $10$ best features, which provide most of the predictive power. Adding additional features generally add no more improvement, and may even decrease the score. The most predictive features are dominated by power spectral features: $\mathrm{PC}_1$ and the power in PSD band  $P_\mathrm{D} = 0.0039-0.031 \,\mathrm{Hz}$ along with a PCA component take the top three spots. Additional improvement is provided by the autoregressive process, with four of its 10 components represented in the reduced feature set, as well as mean and variance of the first hardness ratio, and the skewness of the light curve in the entire energy band. We continue with the rest of the analysis with this reduced set of 10 features, since adding many features does not improve predictiveness. We note that including another $15$ features for a total of $35$ features in the classification does not change the results presented below in any significant way.

\begin{figure*}
\begin{center}
\includegraphics[width=\textwidth]{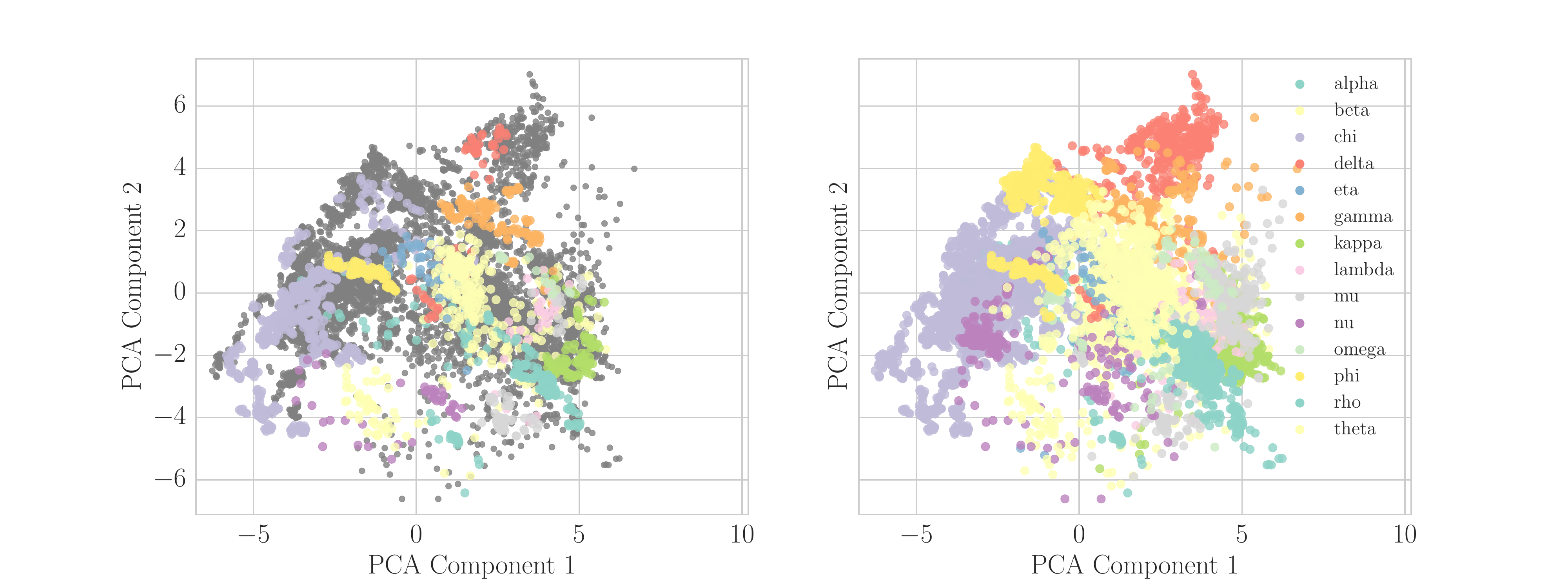}
\caption{Projection of the 10-dimensional feature space into 2 dimensions using PCA. On the left side, the original human classifications in colour and 
unclassified samples in grey. On the right, we show the union of the human classification and the predicted states for the previously unclassified samples. 
Even in this low-dimensional representation, it is possible to see how samples belonging to the same state tend to cluster close together. That this is true also 
for the combined human and machine classified samples indicates that the logistic regression model performed fairly well. We note that seemingly disconnected 
regions are an artifact of reducing 10 dimensions to 2 and plotting many points of different classes in the same Figure.} 
\label{fig:supervised_pca}
\end{center}
\end{figure*}

\section{Supervised Classification}
\label{sec:supervised}

Using the results of the previous sections and the parameters determined in Section \ref{sec:freeparams}, we performed supervised classification using logistic regression on the combined training and validation set for GRS 1915+105 \footnote{For the data set, results as well as all relevant code relevant, see: \url{https://github.com/dhuppenkothen/BlackHoleML}}.
Overall, with a $92.5\%$ accuracy on the test set, the classifier performs very well. This is especially true in light of the small size of the data set as well
as the heavy imbalance between classes, offering only a few test cases for some of the rarer states. At the same time, while the largest class accounts for $\sim 40\%$ of all samples,  incorrect assignments to the dominant state are not the main cause of the mis-classifications, indicating that the model does not underfit by assigning examples of rare states to the more common ones. Additionally, the performance of our model is in line with results from other disciplines, e.g.\ in image classification of dinoflagellates \citep{culverhouse2003}, verb classification in language tasks \citep{merlo2000} and finding humans in images \citep{quinn2010}, where human accuracy is often found to be no better than 
$\sim 90\%$. An illustration of the classification as a whole is presented in Figure \ref{fig:supervised_pca}, 
where we show a 2-dimensional representation of our 10-dimensional features space achieved with PCA. 

\subsection{Confused Classifications}
\label{sec:confusion}

\begin{figure}
\begin{center}
\includegraphics[width=9cm]{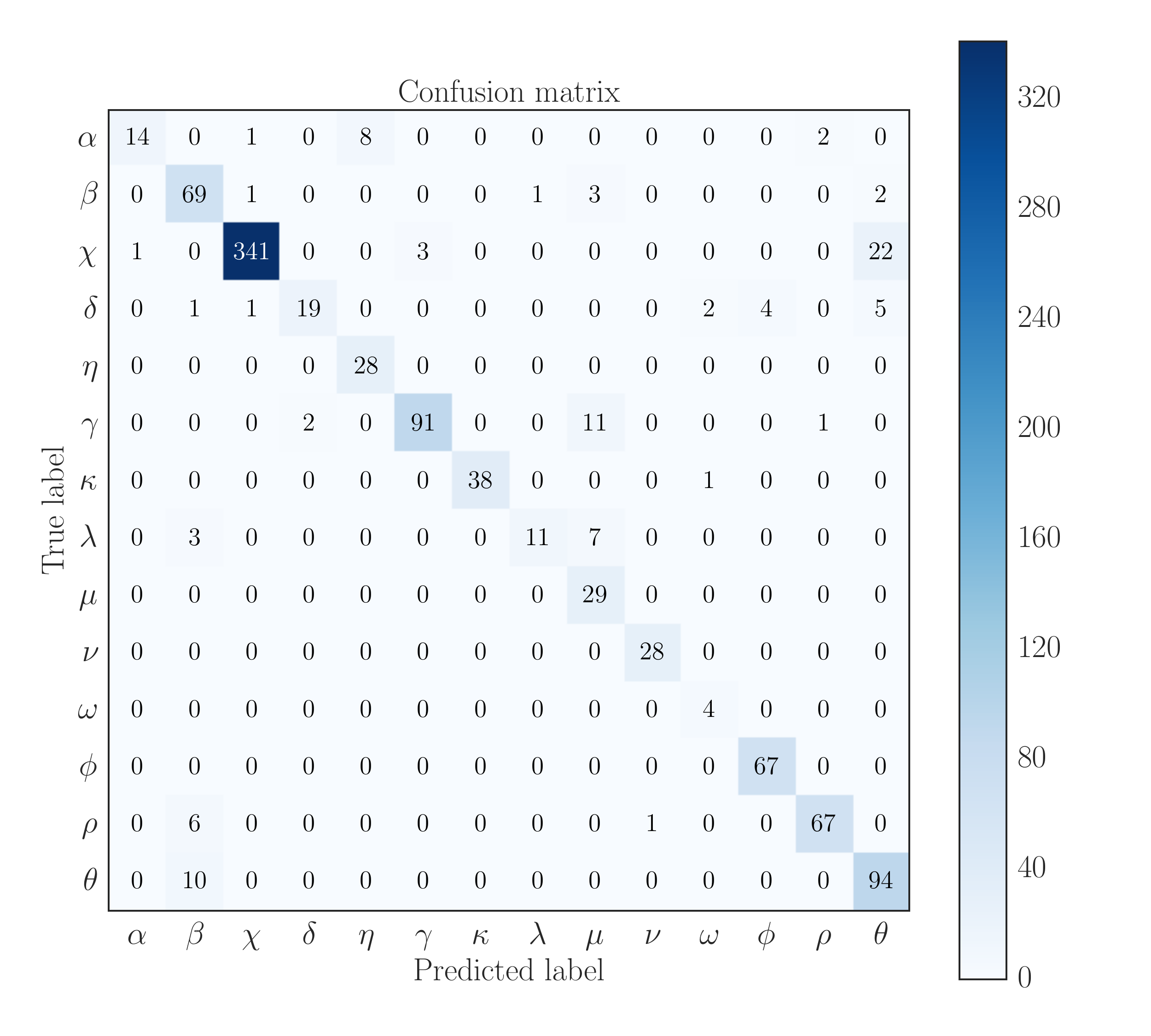}
\caption{Confusion matrix for the machine classification (x-axis) versus the human classification (assumed as the ``true label'') on the 
y-axis. On the diagonal are classes where the human and machine classifications agree. Off-diagonal cases occur where there is a 
disagreement.} 
\label{fig:confusion_matrix}
\end{center}
\end{figure}

In Figure \ref{fig:confusion_matrix}, we show the confusion matrix between the human classification and the machine classification on the 
test set. Generally, the matrix is sparse, and only few classes are confused. For these cases, we visually compared the light curves, hardness ratios and power spectra of 
typical examples (based on the human classification) of both the class chosen by a human and the computer. We find that disagreements between 
human and machine classification fall into one of three categories:
\begin{itemize}
\item{Observations where the particular choice of segment size ($1024\mathrm{s}$) makes it such that only part of the overall pattern is observed in this 
segment. Examples are light curves of the $\alpha$ state, which are occasionally being mis-classified as $\chi$ (when non-flaring) or $\rho$ observations (when flaring). 
The small size (for machine learning purposes) of the data set makes it unfavourable to choose longer segments, thus a small fraction of segments always run the risk of being 
confused in this way. It is worth noting, however, that many of the samples falling in this particular case occur only once or at most twice in the 
test set for a certain combination of classes, thus they are expected to add only a small amount of noise to the classifications.}
\item{For some cases where human and machine classifications disagree, the simple summary statistics and autoregressive model used to represent the variability
 in the light curves fail to fully encode the complexities of the patterns observed in GRS 1915+105. The most striking example is the $\rho$ state, where several 
 segments were instead classified as belonging to the $\beta$ state instead. Looking at the light curve, it is fairly straightforward for the human brain to distinguish 
 both states based on the patterns in the light curve. However, for several cases, the model used for encoding variability was not sufficient to fully appreciate the differences between those two states, in particular since the power spectra look fairly similar. Here, a better model for the light curves would clearly have helped with the classification, however, building such a model for light curves as complex as those observed in GRS 1915+105 is a major undertaking and thus the subject of future work.}
 \item{There are several confused cases where the rigid classification into $14$ states does not capture the behaviour very well. For example, there is a number of examples of the $\chi$ state that have a higher count rate by a factor of $\sim 3$ than typical examples of this state; these light curves are routinely mis-classified as non-flaring parts of either $\nu$ or $\theta$ states, whose dim intervals tend to be much brighter than any typical $\chi$ light curves. Similarly, there is a set of $\gamma$ light curves mis-classified as belonging to the $\rho$ state. This may seem surprising at first; however, closer inspection reveals very regular flares in these light curves as well as a much higher variance than is typically observed in the $\gamma$ state. Perhaps these light curves show a transition between the $\gamma$ and $\rho$ state, and therefore has properties reminiscent of both classes, which in turn confuses the classifier.}
\end{itemize}

The multinomial probability distribution used in the logistic regression model allows for calculating the predicted probability for each class and each sample. We compared the predicted probabilities for each human-generated class and computer-generated class for all of the confused cases and compared them to those cases where human and computer-generated classifications agree. Samples where human and computer agree show a very high predicted probability for the chosen class ($>0.85$ in more than $75\%$ of all cases) and a peaked probability distribution (with low probabilities for all other classes). This is generally not the case for confused cases, which show much flatter probability distributions and the classifier is generally uncertain about its prediction. In these cases, the predicted probability of the class chosen by the computer can be as small as $0.3$ and often close to the probability assigned to the human-generated class. 

\subsection{Overall Distribution of States}

\begin{figure}
\begin{center}
\includegraphics[width=9cm]{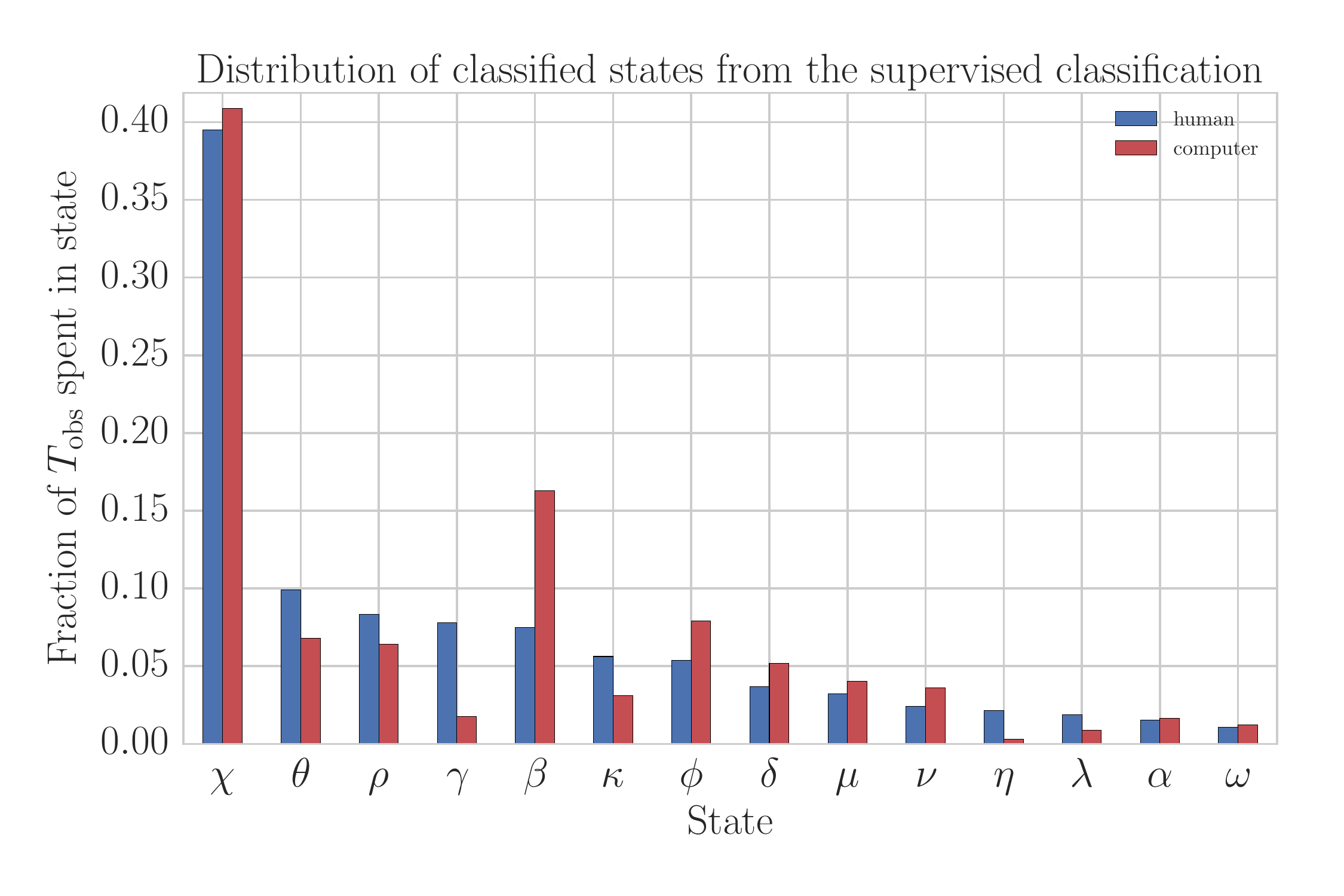}
\caption{Fraction of total observation time $T_\mathrm{obs}$ assigned to a certain state in both the human-classified data (1996-$\sim\!\! 2000$; blue) and the machine-classified data ( $\sim\!\! 2000$ - 2011; red). Durations spent in each state are calculated from the human and computer-generated labels taking into 
account the overlap between segments for long observations.} 
\label{fig:state_durations}
\end{center}
\end{figure}

In Figure \ref{fig:state_durations} we compare the total duration the source spent in each state during the observed intervals for both the human classified part of the data as well as the computer-generated classification. At the same time, this presents a split in time: \citet{belloni2000} and \citet{kleinwolt2002} classified observations between 1996 June and 1999 December, with an additional state identified in an observation on 2003 Mar 6 \citep{hannikainen2003, hannikainen2005}. Trained on these human classifications,
we allowed the computer to find classes for the remaining observations, spanning from $2000$ Jan to the end of \rxte's lifetime in early 2012. 
Assuming that the logistic regression model generally reproduces the human classification, one may then use the data set to search for time evolution in the overall pattern of states. 

We find that broadly, the machine classification reproduces the human classification. Particularly the $\chi$ state remains the most common state to find GRS 1915+105 in. Other states such as $\theta$, $\rho$, $\kappa$, $\mu$, $\alpha$ and $\omega$ are represented similarly often, other classes occur with a significantly different frequency in later observations. It is important to note here that the initial distribution on the state occurrences in the logistic regression model was based on the previous state occurrences, that is, a state with a higher previous occurrence was more probable to occur again than a state that was only seen once or twice. In this context, it is interesting to note the relatively higher fraction of time spent in the $\beta$ and $\phi$ states compared to the human-classified data set.
% AM I feel like prior could be misunderstood here. It's the posterior distribution of \hat{p}(y|x, \theta) based on observing the training data. There is no model p(y), only p(y|x), so I'm not sure if prior is the right word... FIXED

Conversely, the states $\gamma$ and $\eta$ occur much less frequently during later observations compared to the earlier data set. For class $\eta$, this may, to some degree, be due to chance: with only one confirmed observation and the small fraction of telescope time spent on the source, it is intrinsically hard to reliably estimate the duration of the source previously spent in this state. 
% LH: I think this has been commented on recently possibly in Jason Dexter's or Diego's papers? I can check this if you want [CHECK THIS!!!!]
Based on our results from Section \ref{sec:confusion}, it is unlikely that confusions between states play a significant role in explaining all discrepancies between the state durations in the human and computer-classified data sets. Confusions seem to dominate in classes whose fraction of observation time are very similar.

For the classes with the strongest relative discrepancies---$\beta$, $\phi$ and $\gamma$ and $\eta$---we also explored the probabilities of the assigned state in an 
effort to learn how certain the logistic regression model was in its classification for those states. We find that for states $\phi$, $\gamma$ and $\eta$, the classifier is 
fairly certain in its predictions: for example, for class $\phi$, more than $94\%$ of all classified samples have a probability for the source being in state $\phi$ that is 
$>0.8$, and for $98\%$ of all classified samples have a probability of $\phi$ being the true state that is at least twice that of the state with the second-largest 
probability. For this state, there is a small population of samples ($\sim 7\%$) that might be in state $\gamma$ or $\eta$ with a probability of up to $0.4$, that is close to equally likely to the classification as $\phi$. 

For class $\beta$, which shows the largest growth between the early and the late data set, the situation is much less clear. The confusion matrix in Figure \ref{fig:confusion_matrix} shows a significant fraction of other states, most notably $\theta$ and $\rho$ being mis-classified as $\beta$, raising the question whether these mis-classifications could account for the sharp rise in relative time spent in the $\beta$ state in later years. In general, the model is much less certain about these classifications. In only $\sim 60\%$ of all cases, the model predicts a probability for state $\beta$ that is larger than $80\%$, though it still predicts a probability exceeding that of the second-highest class by a factor of two for $83\%$ of all cases. Additionally, while mis-classifications can account for roughly $25\%$ of all observations of this state, the fraction of time spent in state $\beta$ jumped by a factor of $2$. Hence, we conclude that while confusions between states $\beta$, $\theta$ and $\rho$ might account for an appreciable fraction of the observations, they are not sufficient to explain the entire increase in occurrences of this state. 

In summary, we conclude that the effect of an increased number of observations in states $\phi$ and $\beta$ in the machine-classified data are likely real, though the magnitude of the effect for state $\beta$ is hard to guess due to the contamination of the sample of likely mis-classifications.

For states that are much less represented in the machine-classified data set compared to the original human classification, we explore whether these states might have lost samples due to misclassification as well. For this, we found all samples where states $\gamma$ and $\eta$, both of which are almost not present in the 
machine-classified data set, were the second-most probable state, and compared their probability to that of the state the logistic regression model chose for these specific samples. We find that state $\gamma$ comes often second to $\chi$-state observations. However, because the hardness ratios are quite different for both states, we find that the logistic regression model assigns these samples to class $\chi$ with a very high degree of confidence (with a $\chi$-state probability of $>0.8$ in $90\%$ of all samples where $\gamma$ has the second-highest probability). This indicates that the paucity of $\gamma$-state observations in recent years is likely real. Similar reasoning applies to state $\eta$, where we find similar numbers for the confidence that the light curves in question belong to the $\phi$ state instead. Additionally, both $\gamma$ and $\eta$ are extremely rarely confused in the validation and test data sets; in the case of $\eta$, it is more likely to gain false positives from mis-classifications of state $\delta$. 

%For state $\gamma$, on the other hand, the situation is more complex. We find $\sim 200$ segments classified by our model as $\delta$-state observations, of which $\sim 23\%$ have a comparable probability of being state $\gamma$, i.e.\ these could be observations belonging to $\gamma$ that have been misattributed to state $\delta$. However, this fraction is overall not large enough to explain the drop in $\gamma$-state observations between the human- and the machine-classified data set.
Overall, we conclude that there likely was a drop in the occurrence of states $\gamma$ and $\eta$ in the later observations that cannot be explained by confusions with other classes.
% AM concluding sentence? It seems likely that there was an actual drop in phi and gamma? FIXED

\begin{figure}
\begin{center}
\includegraphics[width=9cm]{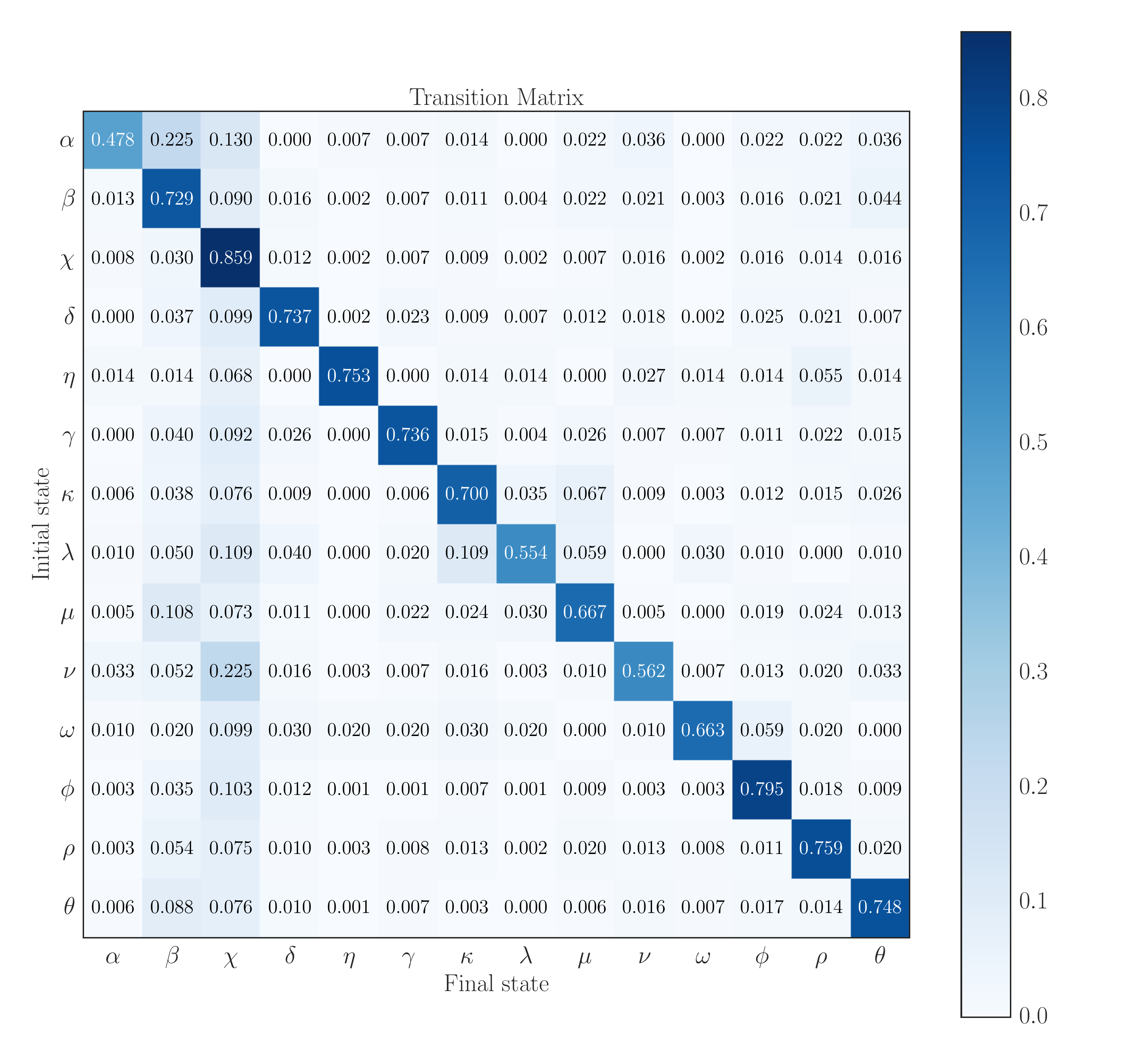}
\caption{Transition matrix of states. We used human labels where available, and labels inferred by the logistic regression model trained on the human labels where 
the latter were unavailable. The matrix presents the probability of arriving in state $x_{t+1}$ given the current state $x_{t}$. The probability is row-wise normalized 
such that the probabilities to arrive in any new state $j$ from a given state $i$ sum to one: $\sum_{j=1}^{N}p(x_{t+1,j} | x_{t,i}) = 1$. The diagonal indicates transitions into the same state.} 
\label{fig:transitionmatrix}
\end{center}
\end{figure}
% AM possibly put numbers in matrix? (I have code). Maybe flip it so that the main diagonal is the identity? (the identity matrix should be everything always stays the same).  [CHECK THIS!]
% AM maybe second figure where you do biclustering to sort by states that have frequent transitions? [CHECK THIS!]

\subsection{Time Evolution of States}

While the logistic regression model employed in the classification task does not include any time dependence, it is instructive to put the classified observations 
into context over the sixteen years of \rxte\ monitoring. In Figure \ref{fig:transitionmatrix}, we show a transition matrix between states. Each row in this matrix 
represents a probability to pass from initial state $i$ to final state $j$, $p(x_{t+1, j} | x_{t, i})$. The transition matrix was constructed by using the human 
classified states for observations where these labels exist, and the computer-based classification for all other observations. We then counted transitions from each 
state $i$ into each other state $j$ for the entire \rxte\ data set, and row-wise normalized such that the probabilities to move into state $j$ from state $i$ sum to one.

Note, however, that there is an important caveat in this procedure: it implicitly assumes continuous observations that are causally connected, that is, the state 
does not change between one observation and the next. This is not true in practice: \rxte\ observed GRS1915+105 for $\sim 2\,\mathrm{ks}$ per day, leaving most 
of the day unobserved. Rapid state transitions are possible, thus the transition matrix here can only be seen as an indication of how state transitions might occur 
in this source. However, a more realistic transition matrix requires more complex (time-dependent) methods that are beyond the scope of this paper.

Overall, it appears that the transition matrix is well-connected: most state transitions are possible, though many occur with a fairly low probability.
Transitions to and from the $\chi$-state occur more frequently than most other transition, which is not surprising given that the source spends the majority of its time 
in this state. Conversely, the probability distribution for leaving the $\chi$-state is fairly flat, indicating that the source is more or less equally likely to go into any 
of the other states.  

There are several other transitions that occur with higher probability. For example, the source is more likely to move from state $\alpha$ into state $\beta$, compared even with $\chi$. 
Some transitions do not occur at all, for example transitions from states $\delta$ and $\gamma$ into state $\alpha$ or from states $\alpha$ and $\eta$ into state $\delta$. In principle, unobserved transitions are of as much interest as those that occur frequently, though their interpretation requires caution. 

\begin{figure*}
\begin{center}
\includegraphics[width=\textwidth]{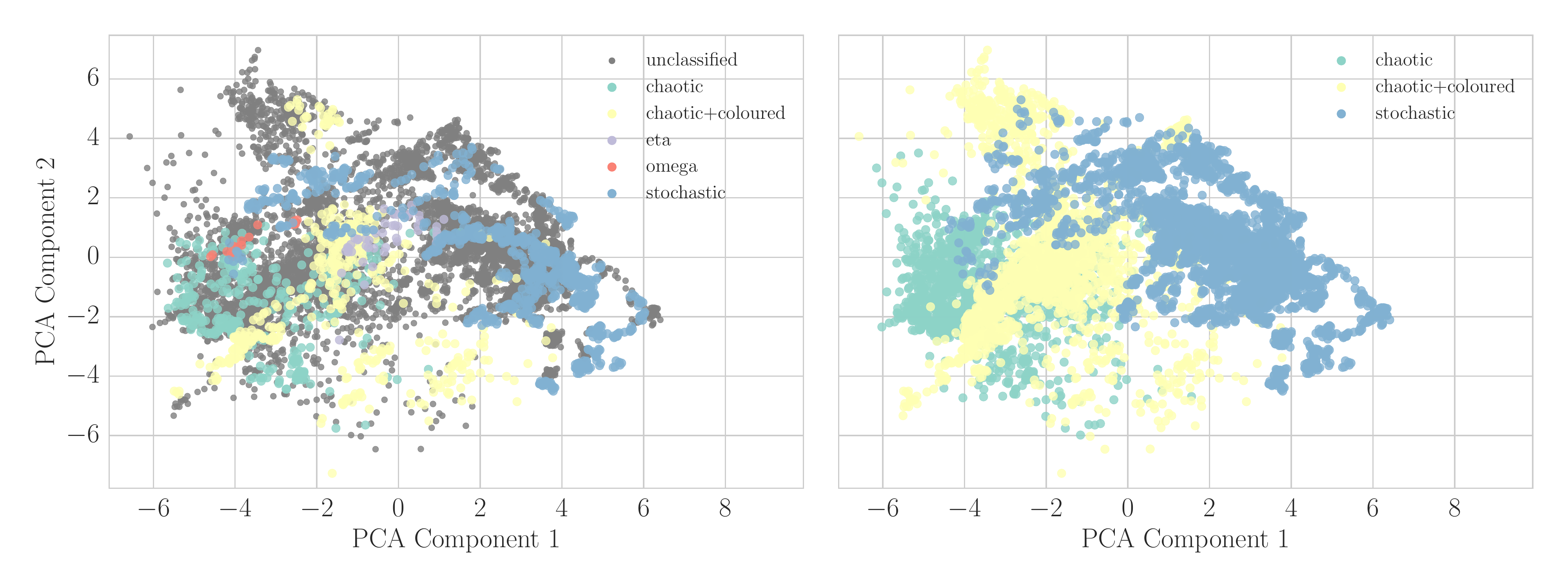}
\caption{PCA representation as in Figure \ref{fig:supervised_pca}, but with the labels following \citet{harikrishnan2011}. In the left panel, we show the human-classified labels in colours, and the unclassified data in grey. We also explicitly mark the samples of classes $\eta$ and $\omega$, for which we have no labels in this scheme. In the right panel, the fully classified data set.} 
\label{fig:pca_physical}
\end{center}
\end{figure*}

While the transition matrix is calculated as a set of probabilities, all we can say about the transitions with a probability of $0$ is that they have not been observed 
during the lifetime of \rxte. This may just as well be due to the lack of continuous observations and the low observational duty cycle as a real physical effects.
In practice, it is interesting to note that while the transition matrix is overall not symmetric (transitions from state $i$ into state $j$ have a different probability from 
transitions from state $j$ into state $i$), there are some notable symmetries.
In particular, transitions between states $\alpha$ and $\delta$, states $\gamma$ and $\eta$, states $\mu$ and $\omega$, and states $\mu$ and $\eta$ never occur in either direction. They indicate that perhaps the transition matrix encodes real physical effects that a better model could capture more efficiently.

\subsection{Supervised Classification with Physically Motivated Labels}

The connection between long-term evolution of the patterns observed in GRS 1915+105 and the underlying physical processes of the 
accrection disc are poorly understood. There is no comprehensive accretion theory that could explain the complex variability observed in the source. 
Therefore, we can only attempt a comprehensive phenomenological description, as done above.
% AM we are left? maybe "we can only attempt" ? FIXED
However, there are attempts to connect the set of 
states with some underlying mechanisms. In particular, \citet{misra2004,misra2006} and \citet{harikrishnan2011} attempted to connect the observed states to 
a non-linear, low-dimensional chaotic system. If true, this would have the advantage of allowing a description of the complex magnetohydrodynamics of the 
accretion disc with a set of ordinary differential equations. They find evidence based on a set of methods optimized for disentangling non-linear dynamics from 
stochastic systems---correlation dimension, correlation entropy and multi-fractal spectra---that some of the classes in GRS 1915+105 show evidence that nearly 
half of the twelve states under consideration exhibit deviations from randomness possibly explained by a non-linear chaotic system. Conversely, other states may 
be well described by stochastic or coloured noise.
% AM would these measures make good features to distinguish the states? [maybe]

In contrast, \citet{polyakov2012} exclusively consider the stochastic components in the light curves using 
flicker-noise spectroscopy (FNS), with the advantage that they can address one of the major shortcomings in the approach chosen by \citet{misra2004,misra2006} and 
\citet{harikrishnan2011}: the presence of Poisson fluctuations, which may contaminate the measures of chaos theory the latter authors use in their analyses.
They find that thirteen of the fourteen states (state $\omega$ had no known observation with sufficient length to perform the analysis) can be classified into 
four phenomenological states based on the characteristics of the stochastic contributions to the light curve: random noise, power-scale variability with a $1/f$ type 
power spectrum, one-scale variability with a single characteristic time scale and two-scale variability with two characteristic time scales. 

In both analyses, while not directly comparable, the fundamental idea is to break down the known states, determined entirely by their patterns in light curves and 
spectral changes, into classes that relate, at least broadly, to underlying physical processes such as stochastic fluctuations of viscosity in the accretion disc or 
changes in the mass accretion rate. Both \citet{harikrishnan2011} and \citet{polyakov2012} point out that their analyses have several drawbacks and shortcomings. 
\citet{harikrishnan2011} does not include either class $\eta$ or class $\omega$ in their analysis, and \citet{polyakov2012} specifically excludes class $\omega$ due to 
a lack of data. For class $\delta$, \citet{polyakov2012} point out that more data is needed to decide whether the chaotic attractors or stochastic fluctuations 
provides a more compelling explanation. These cases imply that the initial classification of the first four years did not provide a sufficient amount of data for the 
different classes.
% AM "either $\eta$ or $\omega$ classes" -> the \eta or \omega classes? FIXED

Here, we apply the classification scheme derived by \citet{harikrishnan2011} as an example of how we can use the machine learning approach developed above 
for a somewhat less phenomenological approach to the data set. 
Following \citet{harikrishnan2011}, we simplify the original labels into three categories: states $\delta$, $\gamma$, $\phi$ and $\chi$ are purely stochastic states
(``stochastic''), whereas $\kappa$, $\lambda$ and $\mu$ show chaotic behaviour contaminated by coloured noise (``chaotic+coloured''), and states $\beta$, $\theta$, $\alpha$, $\mu$ and $\rho$ correspond to to a system showing signatures of deterministic non-linear behaviour (``chaotic'').
We also return previously classified examples of states $\omega$ and $\eta$ to the unclassified data set, since we have no a priori knowledge of their affiliation under this scheme.
\begin{figure}
\begin{center}
\includegraphics[width=9cm]{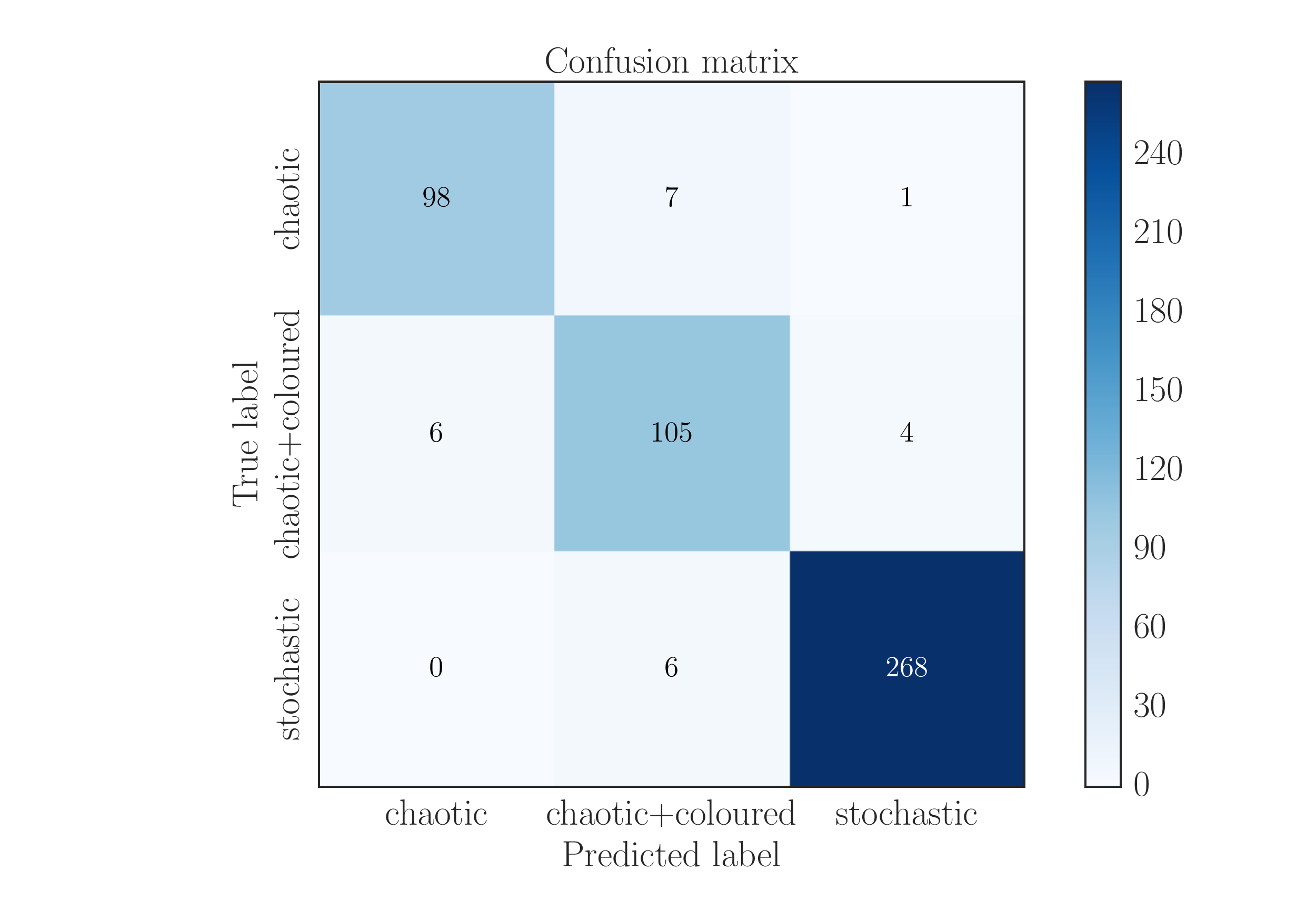}
\caption{Confusion matrix for the physical labels.} 
\label{fig:confusionmatrix_physical}
\end{center}
\end{figure}
% AM definitely numbers in the confusion matrix here. FIXED
% AM the figure label belongs to a different figure partially. FIXED
We then repeat the supervised classification, and find that for this classification problem, the logistic regression model underperforms compared to more complex decision schemes. In particular, random forests provide a higher performance on the validation set ($91\%$ compared to $85\%$ for the logistic regression model). 
% AM: 3-state case? Three classes, each made up of some other states? Also formulation seems odd. FIXED
% AM Silly question: is there code for the papers you reference above. They say they need more data for each state. You have more data now. Can you run polyakov2012 on the machine classified and get classifications for omega or delta? [DONT THINK SO!]
This is not entirely surprising: logistic regression can only draw very simple (linear) decision boundaries in the high-dimensional parameter space, whereas random forests use ensembles of decision trees. Decision trees essentially pose a series of ``if-else'' questions to arrive at a decision for a given sample to belong to a certain class. This allows the decision boundaries between two classes to be much more complex than for the logistic regression model, with the drawback of being much harder to interpret.
% AM random forest maybe not capitalized? FIXED

 For the case with many classes, we have found that the added complexity of the random forests classifier does not lead to an increased accuracy, and conversely the non-linear decision boundaries they draw easily lead to over-fitting. In the classification with $3$ states attempted here, however, we have combined several of the original states into a single new state with a much more complex shape in parameter space (see also Figure \ref{fig:pca_physical}). Therefore, linear decision boundaries result in underfitting, making random forests a more appropriate algorithm for classification here. 

In Figure \ref{fig:pca_physical}, we show the 2-dimensional PCA representation of the samples both before and after classification. The samples of classes $\eta$ and $\omega$, which have not previously been included in this classification scheme, are explicitly marked in the left-hand panel. 
We report a classification accuracy of $91\%$ on the validation set and $93\%$ on the test set. Figure \ref{fig:confusionmatrix_physical} presents the confusion matrix for the classification with the physically motivated labels.
Out of $495$ samples in the combined validation and test sets, only $36$ are confused. $17$ of these confusions occur in the ``stochastic'' state, where $11$ samples are incorrectly 
classified as ``chaotic+coloured'' and $8$ as ``chaotic'', respectively. 
\begin{figure}
\begin{center}
\includegraphics[width=9cm]{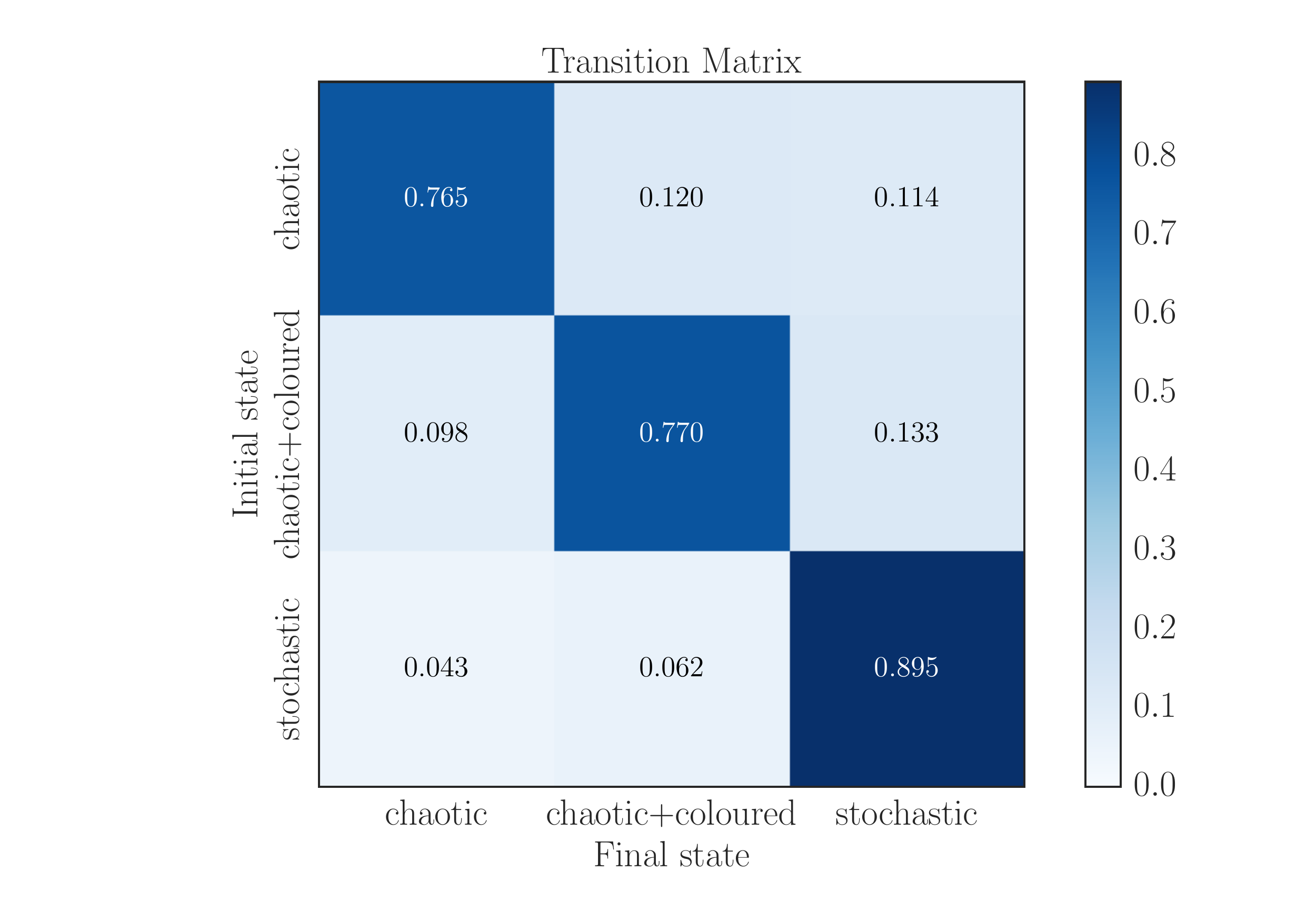}
\caption{Transition matrix for the physical labels.} 
\label{fig:transmat_phys}
\end{center}
\end{figure}

\begin{figure}
\begin{center}
\includegraphics[width=9cm]{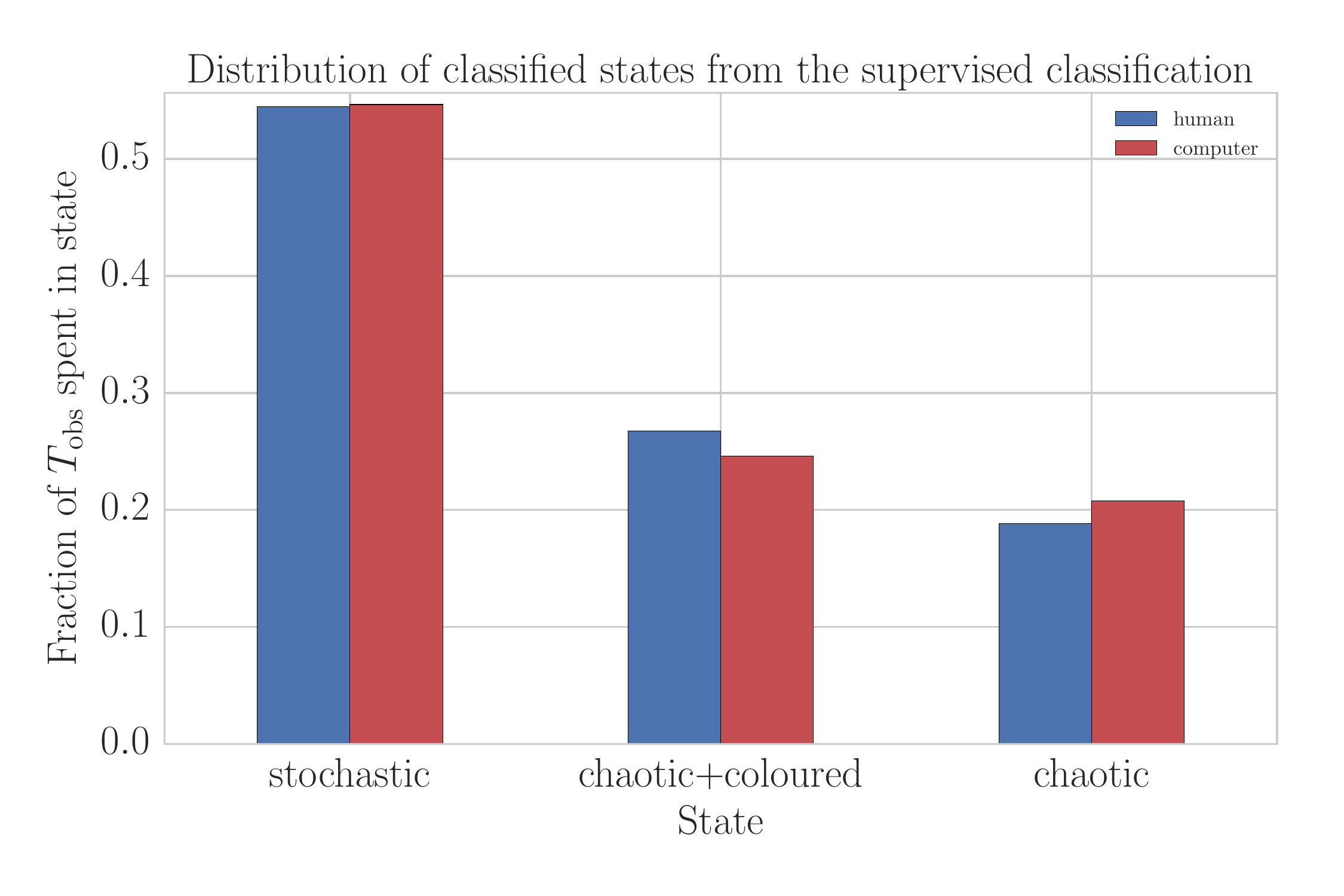}
\caption{Fraction of observed time $T_{\mathrm{obs}}$ spent in each of the three states. In blue, we show the results from the human classification on the first four 
years of data. In red, we show the computer-classified data from the later eight years. The source spends the majority of time showing stochastic variability.} 
% AM maybe rotate labels a bit? Or us hbar? legend very small
\label{fig:duration_phys}
\end{center}
\end{figure}

The transition matrix (see Figure \ref{fig:transmat_phys}) for this classification problem is well-connected. As with the 14-label classification, the source has the 
highest probability to remain in the same state, given the previous state. However, it can easily reach any of the other two states given its current state, with fairly 
similar transition probabilities between $0.043$ and $0.129$. Of course, previously mentioned caveats still apply: this model does not directly encode time-dependence of the states, and we do not know whether there are any visible patterns in how the source transitions between states. Figure \ref{fig:duration_phys} attempts to capture the fraction of observed time $T_\mathrm{obs}$ that the source spends in each state. This is interesting because in principle, it could tell us about the duty cycle of the various accretion regimes and (MHD) instabilities likely responsible for the source's varied behaviour.
% AM not sure I understand the caveats. Why can't this capture time dependent behavior? What is the other caveat? (maybe discuss in person?)
% AM attempts to capture? It is a pretty direct measurement, right? Why attempt?
Since the $\chi$ state in the previous classification with the labels obtained by \citet{belloni2000} is by far the most ubiquitous state, it is unsurprising that more 
than $50\%$ of the time the source can be found exhibiting stochastic variability. The remaining observations are close to evenly split between ``chaotic'' and 
``chaotic+coloured'' states, with the latter being slightly more common.

\begin{figure*}
\begin{center}
\includegraphics[width=\textwidth]{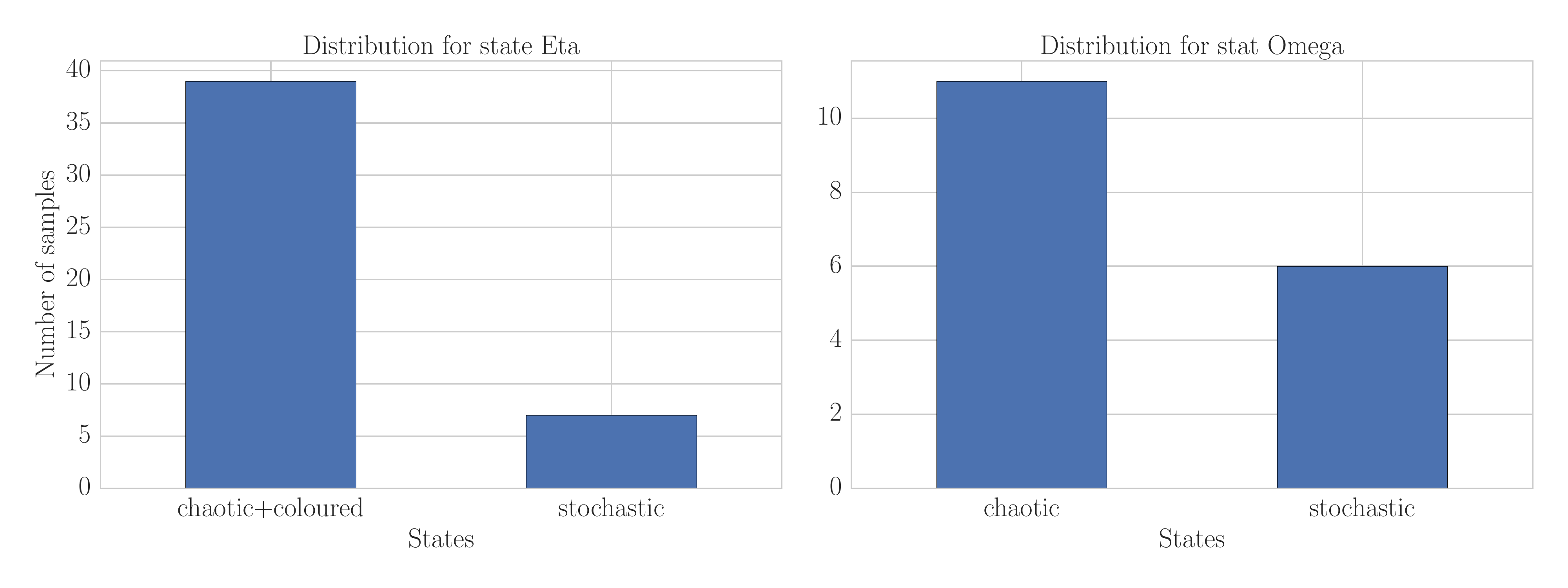}
\caption{Inferred states in the 3-state model for samples classified in the 14-state model as either $\eta$ or $\omega$, which have no identification in \citet{harikrishnan2011}.
We find that most observations in state $\eta$ seem to be closer to other examples of chaotic+coloured variability, while for state $\omega$, the situation is less clear, with a significant 
number of samples identified with some form of either chaotic or stochastic process.} 
\label{fig:etaomega_states}
\end{center}
\end{figure*}

Finally, we also infer the class membership in the physical model of the remaining two states, $\eta$ and $\omega$. 
In Figure \ref{fig:etaomega_states}, we show the distribution of the samples classified by \citet{kleinwolt2002} as $\omega$ and \citet{hannikainen2003} as $\eta$ 
in the classification scheme of \citet{harikrishnan2011}, as inferred from the random forest classifier.
Out of $17$ samples in state $\omega$, $11$  ($65\%$) are classified as ``chaotic'', indicating that this state is perhaps similar to states $\beta$ and $\theta$. This is consistent with the fairly regular pattern of dips observed in \citet{kleinwolt2002} in this state. However, \citet{kleinwolt2002} also point out that in the bright, variable intervals between dips the source shows behaviour reminiscent of state $\gamma$. Segments with no dip or a smaller dip may thus account for the fraction of samples classified as stochastic. This might explain why the random forests classifier has trouble identifying these samples with a single class. It also showcases a general shortcoming in supervised learning: the algorithm will only be trained on what has seen before; if something uniquely new appears, classification is likely to fail.

The situation is much clearer for state $\eta$. A large majority of the samples are classified as ``chaotic+coloured''. This state shows pulses on a 5-minute time scale, but these pulses are overall much less regular than those seen for example in the $\rho$ state. In principle, this may be indicative of a chaotic system driving the processes giving rise to the X-ray emission, perhaps contaminated with stochastic, coloured noise. On the other hand, the hardness ratios are significantly different from all other previously observed classes, one of the reasons why these observations were classified as a new state. 
% AM have you looked at the feature importances of the RF? Would these be interesting to show? Similarly for the many classes, are the logistic regression weights interesting to show? Or what features got selected in the feature selection? (Is there an appendix?)

\section{Discussion and Conclusion}
\label{sec:discussion}

GRS 1915+105 is a remarkable BHXRB. It has been in continuous outburst since 1992, showing at least $14$ different states, compared to at most 
$3$ states in other black hole X-ray binaries.
\rxte's near-continuous monitoring between 1996 and 2011 has resulted in a data set of this source of unprecedented scale and richness, and the existence of its states as well as a subset of previously-classified data makes it an ideal test case for the use of modern machine learning methods in X-ray astronomy.
Here, we classify the entire sixteen-year data set observed with \rxte\ for the first time using a logistic regression model. 
The results allow researchers to pick specific observations where the source inhabited a certain state from the data set for further analyses, vastly improving 
the previous situation, where only a third of the data had known states.

The initial classification was done largely visually: the light curves in the $2-60\,\mathrm{keV}$ band show remarkably complex, but repeating patterns that are 
easily distinguishable by eye (see e.g. Figure (2) in \citealt{belloni2000}). Encoding these patterns in a set of features that a machine learning algorithm can use proves both difficult and instructive.
Many patterns, in particular in states $\theta$, $\lambda$, $\nu$, $\alpha$ and $\beta$ last $\sim 1000\,\mathrm{s}$ or more, similar to the duration of most 
uninterrupted data segments. In most light curve segments we extracted, we see at most one cycle of the pattern, or perhaps only a fragment of it.

A Fourier representation of the data is therefore of limited use here: because of the short duration, it cannot capture the pattern of harmonics generated by the non-sinusoidal nature of the signal. At most, it will be able to capture differences between states at higher frequencies, such as the presence or absence of QPOs. This, however, does not allow us to uniquely distinguish the patterns that are so striking to the naked eye. 

At the same time, each observation in a given state will start at a random phase of the pattern. This immediately makes it impossible to use the light curves directly 
in the machine learning algorithm, since the latter is sensitive to phase shifts. Two light curves of the same state, but shifted in phase appear far 
from each other in feature space. Despite these caveats, our feature engineering in Section \ref{sec:featureselection} has shown that 
the most predictive features are power spectral representations of the data, with the top three features achieving a validation accuracy of roughly $80\%$ by themselves.
This indicates that a better model of the variability might improve the classification further. In contrast, features based on the two hardness ratios had only a minor effect, and only HR1 proved to have a measurable effect on the classification accuracy. In the era of spectral timing, it might be of interest to explore features that tie time and energy closer 
together, such as time lags, covariance spectra and coherence.

There are various strategies that might be successful at improving the features encoding variability. One may use much shorter segments than we have done here, 
which will not encode the full pattern, but parts of it that may be shared across states. For example, the long intervals with a low count rate and low variance in state 
$\alpha$ might be shared with state $\chi$, while the pulses in the same state might be more similar to state $\rho$. The patterns we see would then be repeatable 
cycles of these micro-states. This type of model requires a more complex representation, which is out of the scope of this current work. 
% AM Which is out of the scope of this current work? FIXED

Another strategy is to learn the features from the data itself. This has been a popular approach in a branch of machine learning called ``deep learning'', but generally 
requires vast training data sets with millions of samples. It is unclear how well a deep neural network would be able to learn the structure of the data set from only 
$\sim 8000$ light curves. Alternatively, \textit{autoencoders}, neural networks aimed at learning representations of data, have been used successfully for encoding 
human speech signals for the purpose of both speech recognition and reconstruction \citep[for an overview, see e.g.][]{hinton2012}. 
Speech is similar to the data observed in GRS 1915+105 in the sense that it includes information on vastly different time scales, all of which are important for recognizing the correct word, or in this case, state. These methods can potentially provide powerful encodings of black hole signals beyond GRS 1915+105 and will be explored in future work. 

Another limitation in the approach chosen here arises from the inherent assumption in supervised classification that the examples in the training set are representative 
of the unclassified data, that is, that there are no additional, unrecognized states in the data. If there are states that have so far not been 
recognized in the previously unclassified data set, then the algorithm cannot find them. Unsupervised machine learning methods, which do not make 
use of the human-supplied labels, would be more suitable for this task. A class of models called Hidden Markov Models (HMMs) is one such model which also 
allows for an explicit encoding of the time dependence of the observations. It thus makes it possible to find other states not previously observed, and also infer 
the state the source likely occupied while \rxte\ did not observe it. This is necessary for an accurate inference of the transition matrix, which is limited by the 
low duty cycle of on-target time. Models of this type will be the subject of future work, too.

Finally, it must be mentioned that \rxte\ no longer operates. Another potentially useful avenue of work might be to use transfer learning methods, which allow 
for inference on data sets with a warped feature space compared to the original training data. This is of particular use given the existence of different telescopes 
observing this source, such as \swift\ and \astrosat\, as well as \nicer\ in the near future, which have different sensitivities and energy ranges and will thus create 
a feature space that is similar to that created with \rxte\ data, but not identical. In the future, combining data from different telescopes in the same classification task
could lead to improved insights into the system.
% LM: Mention Astrosat? Could be a nice way to try and get some data(?!) DONE

The exact processes and parameters steering the long-term evolution of GRS 1915+105 are currently unknown. While much attention has focused on 
individual states, in particular the $\rho$ state with its very regular patterns, the long-term evolution of the source, which states it spends its time in and how 
it switches between states has defied explanation. Likely, the observed patterns are due to a complex interdependence between MHD processes in the accretion 
disc and the emission processes producing the observed X-rays. Here, we do not attempt to provide an explanation of the long-term evolution, but instead show 
new ways in which the existing data can be used to derive knowledge about the phenomenology of the source.
% AM Would it be interesting to create a plot of the predicted states over time similar to what was in the paper for the other black hole? Or is the time horizon too long?

\citet{belloni2000} themselves point out that their classification was meant as a phenomenological description only. On the other hand, the observed patterns 
must be tied to the underlying physical processes, in particular the mass accretion rate, thus understanding which states the source spent its time in over the 
past $16$ years plays an important role in understanding how the accretion disc reacts to global changes. Here, we chose a specific model (``stochastic'' versus 
``chaotic'' processes) to highlight the connection of the long-term evolution to real, physical processes in the accretion disc. The idea that the underlying driving 
mechanism could be a chaotic, non-linear dynamical system is compelling, because it reduces the complex problem of magnetohydrodynamics in an 
accretion disc to a system of ordinary differential equations, whereby changes in the disc are driven by changes in global properties such as viscosity or mass 
accretion rate. 

At the moment, no global models of the long-term evolution of GRS 1915+105 exist. However, the feature engineering and classification performed 
in this paper are a first step toward providing the data products that make a comparison between models and data possible. While it is not immediately possible to 
apply the same classifier to other sources, there are other interesting objects for which this approach may be useful. First and foremost, IGR J17091-3624 provides an 
interesting additional test case as it has shown similar states and state changes to GRS 1915+105. Similarly, some Ultraluminous X-ray Sources (ULXs) have shown variability 
similar to $\mu$ and $\kappa$ states of GRS 1915+105, indicating perhaps that ULXs are subject to strongly super-Eddington flows \citep{middleton2011}. 
The methodology presented here could thus be useful to help understand a number of poorly understood objects both Galactic and extra-galactic. 
% LM: Do we need to mention the other possibly similar sources here? 1H0707, possible links to ULXs etc.? Push that this methodology could be useful to help understand a growing number of poorly understood objects both Galactic & extra-galactic? FIXED

\section*{Acknowledgements}

We thank the anonymous referee for their helpful comments.
The authors acknowledge support by the Moore-Sloan Data Science Environment at NYU. The authors thank Brian McFee and Kyunghyun Cho for many useful conversations.

%%%%%%%%%%%%%%%%%%%%%%%%%%%%%%%%%%%%%%%%%%%%%%%%%%



\bibliographystyle{mnras}
\bibliography{grs1915_classification_paper}

%%%%%%%%%%%%%%%%% APPENDICES %%%%%%%%%%%%%%%%%%%%%

% Don't change these lines
\bsp	% typesetting comment
\label{lastpage}
\end{document}